\documentclass[12pt,a4paper]{article}

\usepackage[inner=2.5cm,outer=2.5cm,top=2.5cm,bottom=2.5cm,marginparwidth=5cm]{geometry}

\usepackage[utf8]{inputenc}
\usepackage{amsmath}
\usepackage[table]{xcolor}
\usepackage{xcolor}
\usepackage{multirow}
\definecolor{myOrange}{RGB}{44,14,14}
\usepackage{amsfonts}
\usepackage{afterpage}
\usepackage{amssymb}
\usepackage{array}
\usepackage{tabularx}
\newcolumntype{P}[1]{>{\centering\arraybackslash}p{#1}}
\newcolumntype{M}[1]{>{\centering\arraybackslash}m{#1}}

\usepackage{dsfont}
\usepackage{hyperref}
\usepackage{relsize}
\usepackage{graphicx}
\usepackage{grffile}
\usepackage[section]{placeins}
\usepackage{sectsty}
\usepackage{cancel}
\sectionfont{\fontsize{15}{15}\selectfont}
\subsectionfont{\fontsize{13}{15}\selectfont}
\usepackage{marginnote}
\usepackage[obeyFinal,textsize=tiny,color=yellow]{todonotes}
\usepackage{authblk} 
\usepackage{lmodern,textcomp}
\usepackage[official]{eurosym}
\usepackage{amsthm}

\newtheorem*{prop*}{Prediction}
\usepackage{appendix}

\usepackage{chngcntr}
\usepackage{etoolbox}
\usepackage{lipsum}
\usepackage{adjustbox}
\usepackage{siunitx}
\usepackage{caption}
\newcolumntype{P}[1]{>{\centering\arraybackslash}p{#1}}
\usepackage{enumerate}
\usepackage{float}

\usepackage[style=authoryear-icomp,maxbibnames=9,maxcitenames=2,uniquelist=false, backend=biber, sortcites, natbib]{biblatex}

\defbibenvironment{bibliography}
  {\begin{enumerate}}
  {\end{enumerate}}
  {\item}

\bibliography{library}
\usepackage[english]{babel}
\bibdata{library}

\title{Revealing risk preferences: Evidence from Turkey's 2023 Earthquake}
\author[]{Emily Quiroga\thanks{Corresponding author: emily.quiroga-gomez@uni-hamburg.de}}
\author[]{Michael Tanner\thanks{michael.tanner@uni-hamburg.de}}
\affil[]{University of Hamburg, Germany}
\date{}

\allowdisplaybreaks

\begin{document}

\maketitle

\begin{abstract}
The study on risk preferences and its potential changes amid natural catastrophes has been subject of recent study, yet produced contradictory findings. An often proposed explanation specifically distinguishes between the opposite effect  of realized and unrealized losses on risk preferences. Moreover, higher-order risk   preferences and its relation to post-disaster behaviors remain  unexplored, despite potential theoretical implications. We address these gaps in the literature by conducting experiments with 600 individuals post-Turkey's 2023 catastrophic earthquake, specifically heavily affected individuals who are displaced, those who are not and a control group. Results indicate higher risk-taking in heavily affected individuals when compared to unaffected individuals.  Our results are specifically driven by affected females. We find no pre-existing differences in risk preferences between earthquake and control areas using 2012 data. Within the heavily affected group of individuals, higher house damage—our proxy for realized losses—increases risk aversion, with total destruction of a house inducing even higher aversion. Regarding higher-order risk preferences for individuals heavily affected by the earthquake, we find that prudence is positively associated with self-protective behaviors after the earthquake, specifically internal migration and/or displacement. While precautionary savings shows initially no correlation to prudence, a positive association emerges when considering that prudence is also related to occupational choices, with individuals with stable incomes and who save being more prudent. Our results contribute insights into how disasters influence risk preferences, specifically aiming to address contradictory findings in the literature, while presenting novel evidence on the relationship between prudence and post-natural disaster behaviors.
\end{abstract}

\newpage

\section{Introduction}

Risk and higher-order risk preferences play a crucial role in predicting individuals' decisions  across various domains, including labor market outcomes, health, insurance and medical treatment choices, addictive behaviors, investment decisions, and migration-related choices. The study of these preferences holds fundamental importance for microeconomics and bears numerous practical implications \citep{Schildberg-Horisch2018}.The stability of risk preferences has been of longstanding interest in economic theory \citep{Stigler1977}, with research increasingly addressing this as an empirical question. Natural catastrophes, such as earthquakes, typhoons, tsunamis and hurricanes,  provide a unique opportunity to test for potential changes in risk preferences in the field, given the exogenous nature of such large shocks, which usually manifest in the realm of losses \citep{Hanaoka2018,Ingwersen2023}.

Studies consistently report changes in risk preferences in response to a natural catastrophe, however, there is a contradiction in the literature \citep{Eckel2009, Hanaoka2018, Beine2021, Cassar2017, Li2011}. Approximately 40\% of these papers find that extreme events make individuals more risk-loving, while the remainder find the opposite effect \citep{Abatayo2019}. \citet{Imas2016} offers a potential reconciliation of these contradictory findings. He  states that when individuals experience realized losses, they become less risk-taking, and when these losses are unrealized, the effect is the opposite. Yet, this does not align with the mixed results from the literature on catastrophes. Importantly, no research on catastrophes and risk preferences has explicitly set out to explore the role of realized losses and their magnitude in the context of natural disasters. Particularly, higher-order risk preferences such as prudence, which are linked to self-protective behaviors and precautionary savings, could potentially shape individual responses following a natural catastrophe. However, to  our knowledge there a notable absence of research on high-order risk preferences in the aftermath of natural catastrophes.


We aim to address these research gaps by doing field work with over 600 affected and unaffected individuals in Turkey six to eight weeks after the catastrophic earthquake in Turkey and Syria during 2023. We conducted incentivized experiments, survey-based risk elicitation, alongside surveys on income, asset losses, and other variables of interest. The two earthquakes of 7.8 and 7.6 of magnitude led to the critical damage of an estimated 890,000 living units, with the World Bank assessing  \$34.2 billion in direct damages. Moreover, over 3 million people are estimated to have been internally displaced, according to the International Organization for Migration (IOM)\footnote{https://turkiye.iom.int/earthquake-response}.

To address our first research goal, we test for potential changes in risk preferences between heavily affected individuals and a suitable control group using the global preferences risk preferences module \citep{falk2018}. We select this method as it is previously validated in Turkey, and we gain access to data of the year 2012 to test for potential pre-existing differences between treatment and control groups (as defined by the geography of the earthquake). We conduct experiments with internally displaced individuals in the emergency response centers in the city of Antalya. Our sample also includes  non-internally displaced individuals in the affected cities of Gaziantep, Adana, and Osmanye. The control group comprises individuals from Antalya and its surrounding towns, attempting to exploit the setting as a natural experiment.

Our second research goal aims to disentangle the effect of realized losses, proxied by the magnitude of house damage, on risk preferences of individuals. We conduct incentivized experiments using the \citet{Gneezy} method for risk elicitation, within our sub-sample of affected individuals only, both displaced and not. We also explore other potential proxies of realized losses that go beyond a strict understanding of material losses stemming from house destruction.

Our third goal entails assessing relation of higher-order risk preferences, specifically prudence, on post earthquake behaviours related to precautionary savings and self-protective behavior\footnote{i.e., primary prevention, explained as behaviors/decisions reducing the likelihood of a loss occurring with the loss size being exogenous\citep{Eeckhoudt2005}.}. We conduct incentivized experiments with the subsample of affected individuals, both displaced and not. We utilize the prudence experiment developed by \citet{Eeckhoudt2006}, using the modified version of \citet{Schaap2021} for developing settings. We propose that internal displacement and/or internal migration can be understood as a form of self-protective behavior, and likewise, we test for precautionary savings after the earthquake.

Our findings align with expectations, revealing that individuals heavily affected by the earthquake display different risk-taking tendencies compared to our control group. Specifically, individuals affected by the earthquake are significantly more risk taking when compared to the control group. Additionally, our analysis using 2012 data from \citet{falk2018} indicates no pre-existing differences in risk preferences between individuals from the earthquake and control areas. This further supports that observed differences can be attributable to the earthquakes effect. Gender-wise, we observe potential heterogeneous impacts, with our overall results primarily driven by changes in females' risk preferences. Evidence suggests that a negative income shock might underlie these gender-specific changes. For our second hypothesis, we find a positive association between the level of house damage and increased risk aversion among heavily affected individuals, consistent with \citet{Imas2016}. Moreover, the magnitude of realized loss matters, with total house destruction leading to higher risk aversion compared to lower levels of damage. Our results capture the broader contradiction evidenced in the literature, where the earthquake resulted in overall increased risk-taking compared to the control but induced heightened risk aversion within the affected group, especially for those experiencing catastrophic losses. 

For our third research goal, we find that prudence has a significant and positive association with self protective behaviour, i.e the decision to migrate in our setting. For our exploration of precautionary savings and prudence, we find no initial correlation. However, considering that prudence also plays a role in occupational choice, i.e., prudent individuals might prefer less risky income paths \citep{Fuchs-Schundeln2005}, we control for occupational self-selection and find a positive association between savings and stable incomes and prudence. 

Overall, our results are robust to various checks and regression specifications, including analyses with matched and unmatched samples. Our findings are align with previous research in different strands of the relevant literature, such as the stability of risk preferences, the effects of natural disasters, the impact of prior losses on risk attitudes, and empirical studies of prudence and related behaviors.

The remainder of the paper is organized as follows. Section (2) provides a brief literature review. Section (3) details our hypothesis, the field setting and our experimental design. Section (4) describes the data. Section (5) presents our analysis. Section (6) discusses  results and concludes.

\section{Risk Preferences and Natural Disasters}
\label{section: field setting}

In recent years, economists increasingly delve into the examination of the stability of risk preferences, and the body of evidence on this subject is expanding rapidly. This heightened interest is, in part, a response to the recognition that the stability of preferences is, to some extent, an empirical question \citep{Schildberg-Horisch2018}. Moreover, it challenges the long-standing assumption in economics that individual risk preferences remain constant over time, a concept argued by \citet{Stigler1977}. Examining the constancy of (risk) preferences is a fundamental aspect of microeconomics with significant practical implications. This investigation is crucial because an individual's inclination towards risk-taking can predict various aspects of labor market performance, health outcomes, addictive behaviors, investment choices, and decisions related to  migration \citep{Schildberg-Horisch2018}. 

In existing empirical evidence, negative shocks such as financial crises and natural catastrophes could have persistent effects on risk preferences \citep{Hanaoka2018}. However, there is no consensus regarding the direction of the impact of extreme event, and little understanding about mechanisms through which these events change risk preferences \citep{Abatayo2019}. For instance, \citet{Eckel2009} elicited risk preferences in a sample of hurricane Katrina evacuees twenty days after the hurricane and compare it with a sample of the same evacuees and non-evacuees ten months after the hurricane. They found that women were significantly more risk loving for the sample evacuees 20 days after the hurricane. \citet{Hanaoka2018} also studied whether risk preferences changed after the 2011 Great East Japan Earthquake. They used a representative survey that follows risk preferences on individuals before, and one and five years after the earthquake. They found that men who experienced a higher intensity of the earthquake become more risk tolerant.  \citet{Beine2021} conducted a survey and a field experiment in Tirana, Albania before and (coincidentally) after two major earthquakes hit Albania during 2019.
They found that there is a significant increase in risk aversion after the earthquakes. \citet{Abatayo2019} compared individuals from communities in the Philippines that were directly hit by a typhoon with those that were not, observing evidence that those affected by the typhoon are less risk averse. Additionally, they found  strong evidence that females affected by the typhoon are more risk-loving than females unaffected by the typhoon.  \citet{Ingwersen2023} found that survivors of the Indian Ocean tsunami from Indonesia who were directly exposed to the tsunami made choices consistent with greater willingness to take on risk relative to those not directly exposed to the tsunami. Yet these differences were short-lived,  a year later, there is no evidence of differences in willingness to take on risk between the two groups.

In a survey of the existing literature investigating the effect of extreme events on risk preferences \citet{Abatayo2019} showed that approximately  40\% of these papers found that extreme events make individuals more risk-loving and the remainder found the opposite effect. \citet{Imas2016} proposed a potential reconciliation of these contradictory findings, arguing that individuals with \textit{realized} losses, i.e. those that are experienced, take on less risk. On the contrary, individuals with \textit{paper} losses, i.e., holding a losing stock or not cashing out after a loss, take greater risk.  \citet{Imas2016} defined \textit{realization} as an event in which money or another medium of value is transferred between accounts, where these accounts could be real, (e.g., broker- age, savings), or mental accounts. If losses stemming from a natural disaster are preponderantly realized losses, then relative risk aversion would dominate across the findings in the literature. However, the mixed evidence suggests that extreme events have inconsistent effects on risk preferences, hence timing and context matter, specifically in developing settings \citep{Abatayo2019}.

In addition to risk preferences, some decisions also depend on higher order risk attitudes. Second order risk aversion leads individuals to opt for higher levels of prevention in the context of self-insurance. However, when it comes to self-protection, relying solely on risk aversion is insufficient to determine the optimal level of preventive effort.  Third-order risk aversion, i.e. prudence, also affects the optimal level of prevention \citep{Eeckhoudt2005}. The effect of prudence on preventive effort has been mostly approached theoretically, with conflicting arguments if prudence leads to more or less preventive effort \citep{Eeckhoudt2005,RePEc:eee:matsoc:v:58:y:2009:i:3:p:393-397}. 

Third-order risk aversion affects precautionary saving due to changes in the distribution of a future income stream  which are determined by individuals’ prudence \citep{Eeckhoudt2006}. The degree of prudence individuals exhibit has implications on a wide range of economic applications, from bargaining, rent seeking, behaviour in risky occupations to health related decisions. For example, \citet{Felder2014} showed that prudent individuals test and treat earlier in the health domain. Moreover the degree of prudence on experimental measures is predictive for  wealth, saving, and borrowing behavior \citep{Noussair2014}. Additionally, prudence preferences play a vital role in shaping decisions regarding preventive behavior. \citet{Eeckhoudt2005}, along with \citet{Experiments2008}, examined the impact of prudence preferences on preventive measures. They clarified the distinction between two types of prevention: (1) self-protection (primary prevention), which reduces the likelihood of a loss occurring (with the loss size being exogenous). (2) self-insurance (secondary prevention), which focuses on minimizing the magnitude of a loss (while the likelihood of occurrence is exogenous) \citep{RePEc:ucp:jpolec:v:80:y:1972:i:4:p:623-48}.

In light of the existing literature, our aim is to contribute specifically to the research on the effect of catastrophes on risk preferences, particularly focusing on the role of the magnitude of realized losses in influencing risk preferences. Additionally, we seek to explore prudence-related behaviors that might theoretically be linked to post-natural disaster responses, an area lacking in empirical evidence.

\section{Hypothesis, Setting and Experimental Design}

\subsection{Hypothesis}
 
In this section we present our hypotheses, which we derive from the literature and inform the experimental design and data analyses.

First, in line with the literature researching natural catastrophes and changes in risk preferences \citep{Hanaoka2018,Ingwersen2023, Abatayo2019} we expect:

\begin{prop*}
	\label{compare}
	
	\begin{enumerate}
		\item[1:] The impact of the earthquake leads to changes on the risk preferences of individuals who were severely impacted as compared to our control group
        
	\end{enumerate}
 \label{H1}
\end{prop*}

For the first hypothesis, we expect significant changes in risk preferences. However, given the conflicting results across the literature, we are agnostic about the direction of the changes. In light of those conflicting results, we set out specifically to test in the field the potential explanation that asset losses, in the form of realized assets as proposed by \citet{Imas2016}, might be a driver of results towards increased loss aversion. Therefore, for the second hypothesis we set out to look into risk preferences within the treated group (affected by earthquake) and the role of asset losses. These losses are understood as varying degrees of house damage of the respondents, as a proxy for realized losses. Therefore we propose:

\begin{prop*}
	\label{compare}
	\begin{enumerate}
		\item[2:] Within individuals affected by the earthquake, realized losses proxied by house damage is correlated to increased risk aversion
        
	\end{enumerate}
 \label{H2}
\end{prop*}

Finally, we address the relevance of  higher order risk preferences, i.e. prudence, in our study setting from two potential angles. First, given the literature on precautionary savings, we aim to explore the differentiated effects on savings after the earthquake for heavily affected individuals. Second, we explore the association of self protective behaviour/primary prevention and prudence in our setting. We propose internal displacement is at least partially a mechanism to reduced the likelihood of a loss occurring (with the loss size being exogenous), and therefore we expect an association to prudence. For the latter, we are agnostic regarding the direction of the effect, as there are conflicting theoretical predictions that either lead to positive or negative association. Therefore we propose:        

\begin{prop*}
	\label{compare}
	Within individuals affected by the earthquake, higher order risk preferences are correlated to
	\begin{enumerate}
		\item[3:] post earthquake saving behaviour, as a form of precautionary savings, thus prudence should have a positive correlation with saving behaviour
  \item[4:]  Internal displacement, as a form of primary prevention/protective behaviour.
	\end{enumerate}
 \label{H3}
\end{prop*}

\subsection{ Field setting}

On February 6, 2023, an  earthquake of 7.8 (Richter Scale) struck southern and central Turkey, as well as northern and western Syria, marking the strongest seismic event in Turkey in over 80 years. Approximately 9 hours later a second earthquake with a 7.6 magnitude occurred to the north-northeast in Kahramanmaraş province. By the 20th of March of 2023, the total death toll of over 57,000 (50,000 in Turkey and 7,000 in Syria) \citep{Hussain2023}

The 2023 earthquakes originated from the East Anatolian Fault (EAF), a major tectonic structure in the eastern Mediterranean, separating the Arabian and Anatolian tectonic plates. Only in Turkey, over 13 million people experienced moderate to high levels of ground shaking in a region already grappling with a high number of COVID-19-related illnesses \citep{DalZilio2023}, with over 9.1 million being directly affected\footnote{https://reliefweb.int/report/turkiye/kahramanmaras-earthquakes-turkiye-and-syria-31-may-2023}.

Over 1.5 million people lived below the national poverty line in the affected provinces according to 2021 data. Moreover, 52\% of homes in these provinces were constructed after 2001 when strict new building regulations were enforced following the destructive 1999 magnitude-7.4 Izmit earthquake \citep{Hussain2023}. Despite these regulations, over 230,000 buildings were damaged or destroyed. Later estimations approximate the number of destroyed or critically damaged units to be 890,000, with more than 1.8 million units being lightly damaged. However, those with light damage do not necessarily provide adequate living conditions\footnote{ United Nations Office for the Coordination of Humanitarian Affairs (OCHA)}. 

Economic damages are estimated at \$34.2 billion in direct damages in Turkey from the earthquakes, according to the World Bank’s \citeyear{WorldBn2023}  Global Rapid Post-Disaster Damage Estimation (GRADE) Report \citep{Gunasekera2023}. Additionally, the most-affected Turkish provinces hit by the 2023 earthquakes suffer from higher levels of poverty compared to western Turkey. Internal displacement is estimated at 3 million people, according to the International Organization for Migration (IOM) \footnote{https://reliefweb.int/report/turkiye/turkiye-2023-earthquakes-situation-report-no-11-23-march-2023-entr}. As a response, the government implemented  distribution centers of goods for the earthquake victims to ensure access to critical goods and services. These centres were located in the main cities, where most of the victims were displaced.

\subsection{Experimental Design}
\label{section: experimental design}

\begin{figure}[ht!]
  \centering

  \includegraphics[scale=0.5]{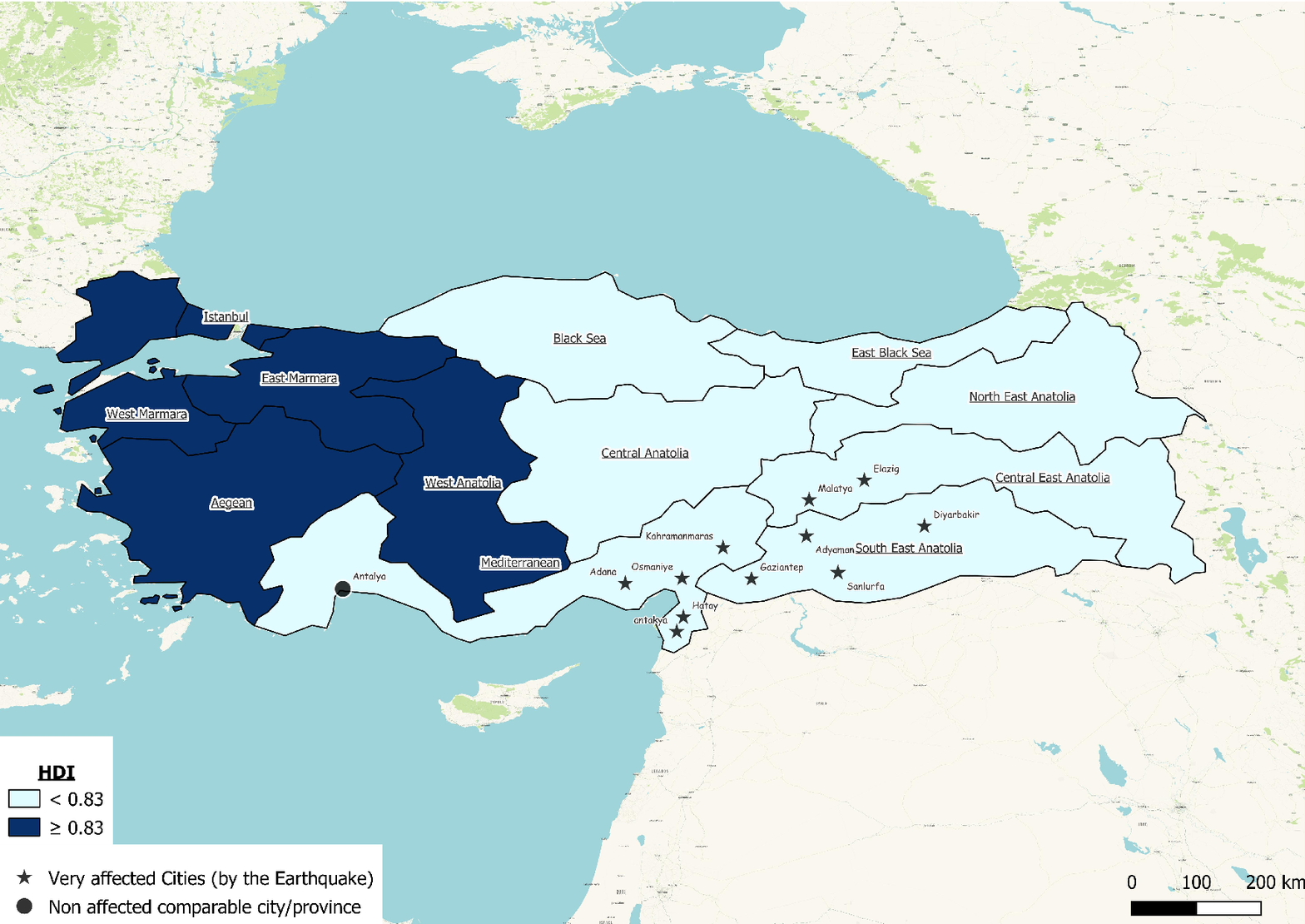}
  \caption[Map of Human Development Index (HDI) of Turkey in 2021]{Map of Subnational Human Development Index (HDI) as of 2021 per Turkey's NUTs1 Statistical Regions with geographical location heavily affected cities and our selected control city. The Mediterranean region is located in the southern part and includes the province of Antalya and the affected cities.}
  \label{Fig:HDIMap}
\end{figure}

In this section we describe the research design for each hypothesis. Our first hypothesis aims to detect potential changes in risk preferences which are attributable to the earthquake. For this, we determine our treatment area and relevant control area, ensuring that the latter is of comparable nature to the affected regions. 
We first define treatment as individuals from the geographical area heavily affected by the earthquake, the allocation can be expected to be as good as random. To ensure an appropriate control area, we analyse unaffected geographical areas which could serve as plausible comparison group. We  identify a region with similar economic and human development indicators as the affected region, but geographically distant from the epicenter of the earthquakes. We conducted this  analyses looking at 12 statistical regions in Turkey and their specific Human Development Index (HDI). Figure \ref{Fig:HDIMap} presents 12 regions in Turkey, with regions below the national mean of HDI in light coloring, and the regions with HDI above the national mean in darker. Cities which are defined as heavily affected are marked. Differences in HDI between the east and west of the country makes most of the western provinces not ideal controls. The Mediterranean Region is unique with affected and unaffected cities due to geographical distance, whilst also having an HDI which is comparable to all of eastern Turkey. We select the province of Antalya as potential control for the experimental design of hypothesis one. Table \ref{tab:gdpPerCapita} presents the GDP per capita per province for both the most affected cities and our selected control province (in bold). The table also shows other provinces that were not affected and served as receivers for internally displaced/migrants (marked with a *), yet deemed unsuitable as a control region.

\begin{table}[!ht]
    \centering
    \begin{tabular}{@{}p{3cm}p{3cm}p{3cm}@{}}
        \hline
        \hline \\[-1.8ex]
        Province & 2020 & 2021 \\
        \hline
        Adana  & 6,291 & 6,977 \\
        Ankara* & 12,041 & 13,020\\
        \textbf{Antalya}  & 7,252 & 8,983\\
        Elazig & 5,944 & 6,272\\
        Gaziantep & 6,763 & 7,819\\
        Hatay  & 5,385 & 6,785 \\
        İstanbul* & 13,931 & 15,666\\
        İzmir* & 9,967 & 11,668\\
        Kahramanmaraş & 5,601& 5,997\\
        Malatya & 5,147 & 5,355\\
        Osmaniye & 5,133& 6,256\\
        \hline
        \hline \\[-1.8ex]
    \end{tabular} 
    \caption[Gross Domestic Product per Capita at province level]{Gross Domestic Product (GDP USD) per Capita at the province level in Turkey. Selected control province in bold. * Unaffected provinces deemed unsuitable control areas.}
         \label{tab:gdpPerCapita}
\end{table}

Some victims of the earthquake were internally displaced from their homes in the heavily affected regions. Most of them established themselves in temporary housing and camps on cities not affected by the earthquake. These individuals  visited the centers where the government provided food and clothes for free to all affected. We perform most of our surveys and experiments in the centers located in the city of Antalya. In these centers the participants were waiting to receive their goods, time in which we carried out the survey elicitation of risk preferences. Beside the surveys of earthquake victims performed in Antalya, we also execute the study in situ in the regions of Gaziantep, Adana and Osmaniye. Therefore, we ensure to have a population of both, affected individuals internally displaced and not displaced in our sample. For our control individuals we surveyed individuals in the city of Antalya and surrounding areas, as these also serve as comparison given our priority for timely field work and the existing circumstances on the ground\footnote{Movement to the Eastern regions was severely complicated due to the immediate effects of the earthquake, and travelling to the southern regions advised against due to security concerns.}. The risk elicitation surveys and experiments were carried out starting the 24th of March, to ensure capturing the effects of the earthquake as much as possible. Ethics approval for our design was provided by the University of Hamburg, additionally informed consent was obtained from participants.

We assess individual risk preferences for treated and control individuals using the risk module survey designed by \citep{falk2018} to test our first hypothesis. The Global Preference Survey (GPS) designed by \cite{falk2018} elicits economic preferences from 80.000 people in 76 countries, including Turkey. The measures include time preference, risk preferences, positive and negative reciprocity, altruism and trust. The risk preference module includes a lottery choice using the staircase method and a self assessment question about the willingness to take risks (See Annex \ref{sec:riskTreeFalk}). The results of the lottery stair case and the self assessment question were converted into a single \textit{risk index} using the method and weights described in \citet{falk2018}. Aside from its robustness and wide application in developing settings, we select this method as we gained access to data by \cite{falk2018} on Turkey prior the earthquake. Employing the identical risk elicitation method allows us to conduct checks to explore pre-existing differences in risk preferences between individuals from our treatment and control areas. All individuals also completed an accompanying survey designed to reveal loss of assets/Level of house damage, existing informal support networks, income changes and other variables of interest.

To test the second hypothesis we conduct incentivized experiments with individuals who were heavily affected by the earthquake, thus only the treated group from hypothesis 1. We conduct these experiments with both internally displaced individuals, and also in situ in heavily affected areas\footnote{Given field constraints only individuals from Adana and Osmaniye conducted incentivized experiments, individuals Gaziantep only conducted survey module by \cite{falk2018} }. To elicit the the magnitude of realized losses our survey  inquires levels of house damage, for individuals affected by the earthquake, loss of income, and existing networks. For our incentivized experiments we use the \citet{Gneezy} experiment in risk taking. This experiment mimics an investment decision, where participants receive an endowment and  decide which part of it is invested into a risky asset. This asset pays off three times the investment with a probability of 50\%, and zero otherwise. The level of risk aversion in individuals is determined by their investment choices; higher investments reflect lower levels of risk aversion. The figure \ref{fig:riskWilligness} shows the structure of the experiment. The relative simplicity of the method, combined with the fact that it can be implemented in a single trial and basic experimental tools, makes it a useful instrument for assessing risk preferences in the field \citep{Charness2013}. Moreover, when compared to other experimental methods, the \citet{Gneezy} approach is more consistent in its findings, specifically in developing/rural settings \citep{Charness2016}, which is a matter of importance for the design.
 
 We also design and implement a survey  to elicit a measure of perception of asset loss, specifically information on the level of house damage by the earthquake. Within this survey we also gather data regarding income, savings, married status, education, migration and mathematical abilities among others. 

For our third hypothesis, we implement a second incentivized experiment with the same group as hypothesis two. This experiment aims to assess  higher order risk attitudes, specifically prudence, with the individuals affected by the earthquake. We implement the prudence experiment developed by \citet{Eeckhoudt2006} using the modified version for developing settings described by \citet{Schaap2021}. To assess participants' prudence, they were presented with 5 binary choices. These choices involved assigning a risk with a mean of zero to either the high or low outcome of a lottery. The lottery and the mean-zero risk were determined by independent coin-flips, represented in figure \ref{fig:EeckhoudtSchlesinger}. Allocating the mean-zero risk to the high (low) outcome indicated prudent (imprudent) preferences \citep{Eeckhoudt2006}. The specific choices are detailed in Table \ref{table:PrudenceChoices}. Participants considered one choice at a time and received no feedback until the end of the session. Only one of the choices was paid out at the end. The first choice is referred to as the baseline choice, in the subsequent four choices, the expected payout for opting either option A or option B in the baseline choice was increased. This adjustment created an incentive for choosing the imprudent or prudent option, respectively.

These choices are identified by the expected payout of the prudent option compared to that of the imprudent option. For instance, in the choice labeled ``+10," the expected payout of the prudent option is ten lira higher than that of the imprudent option. This framework enables the detection of inconsistent behavior concerning payout maximization. For instance, if a participant initially chose the prudent option in the baseline but later opts for the imprudent option in the ``+10" choice, this decision is considered inconsistent. Prudence is quantified based on the number of prudent choices made out of the five available options \citep{Schaap2021}.

\begin{table}[!htbp] \centering 
\begin{tabular}{@{\extracolsep{5pt}}lcccc } 
\\[-1.8ex]\hline 
\hline \\[-1.8ex] 
 & \textbf{Option A} & \textbf{Option B} & \textbf{Frequency A} & \textbf{Frequency B}\\
\cline{2-5}
Baseline & $60|(+30 |- 30)||40$ & $60||40 (+30 |- 30)$ & 208 & 90\\
\\[-0.9em]
Prudence +10 & $70|(+30 |- 30)||50$ & $60||40 (+30 |- 30)$ & 210 & 88\\
\\[-0.9em]
Prudence -10 & $60|(+30 |- 30)||40$ & $70||50 (+30 |- 30)$ & 128 & 170\\
\\[-0.9em]
Prudence +20 & $80|(+30 |- 30)||60$ & $60||40 (+30 |- 30)$ & 222 & 76\\
\\[-0.9em]
Prudence -20 & $60|(+30 |- 30)||40$ & $80||60 (+30 |- 30)$ & 100 & 198\\
\\[-0.9em]
\hline 
\hline \\[-1.8ex] 
\end{tabular} 
  \caption{Prudence experiment lottery } 
    \label{table:PrudenceChoices} 
\end{table} 

Table \ref{table:PrudenceChoices} presents the list of lotteries with the respective frequency of individuals who choose the respective option. Each choice has an option A or B and is composed with two steps. For instance, in the first step of the option A at the baseline, if the coin falls heads the individual gets 40 Lira otherwise they get 60 lira and proceed to the second step. Then the coin is flipped again, if the coin falls tails the individual will receive 60 lira plus 30 lira (total of 90 lira), otherwise they get 60 lira minus 30 lira (total of 30 lira). The corresponding notation of this procedure is ``$60|(+30 |- 30)||40$". The same reasoning follows the rest of the choices (See figure \ref{fig:EeckhoudtSchlesinger})

\section{Data}

\begin{figure}[ht!]
  \centering
  \includegraphics[scale=0.5]{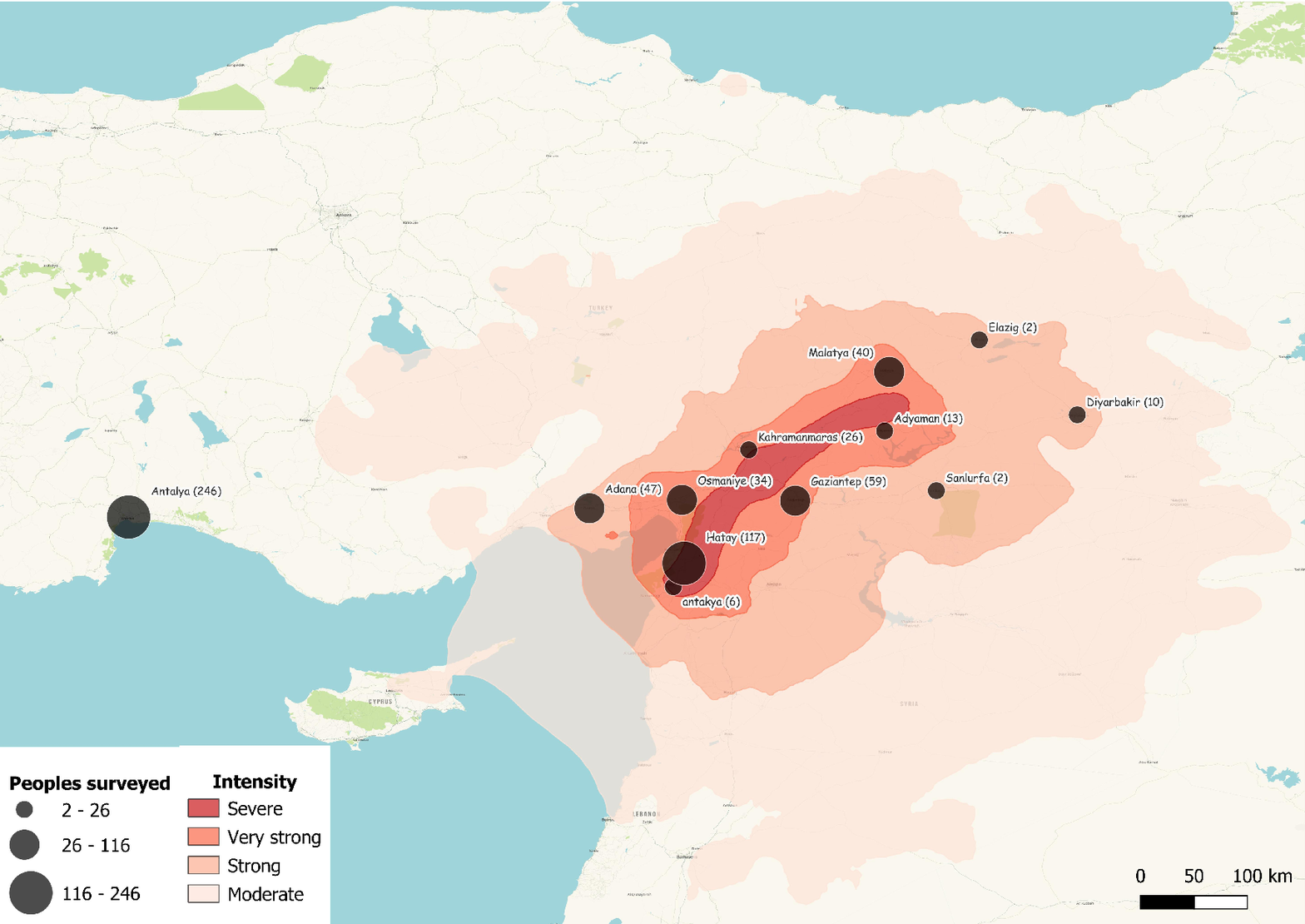}
  \caption[Map showcasing sample distribution and earthquake intensity]{Map showcasing sample distribution and earthquake intensity using U.S Geological Survey Data\footnote{https://earthquake.usgs.gov/earthquakes/eventpage/us6000jllz/shakemap/intensity} and scale based on \citet{Worden2012} }
  \label{fig:mapStudy}
\end{figure}

In this section we present  a description of the collected data prior to analyses. Figure \ref{fig:mapStudy} shows the treated area of our study (determined by the geographical location of the earthquake), and plots all surveyed individuals \textit{N=602} (control, treated internally displaced and treated in situ). By randomly surveying heavily affected individuals at the goods collection center we are able to get a representative sample of the whole affected zone as show in Figure \ref{fig:mapStudy}. Our sample holds individuals from all the most affected cities, with the circles on the figure representing the number of individuals surveyed and the city where they originally came from. 

Table \ref{tab:summaryStatistics} shows the characteristics of the population surveyed. In total we collect data of 602 participants. To test the first hypothesis we use the non-incentivized survey based  sample that aims to measure risk preferences following the procedure described in \citep{falk2018}, compring of all individuals. For the second and third hypotheses we use the sample composed of individuals who participated in incentivized experiments (302 participants). This sample is a subset of the ``treatment group" (Third column in table \ref{tab:summaryStatistics}). The variable \textit{Savings} takes the value of 1 if the person has savings to support themselves for the next six months or zero otherwise. \textit{Math} is a self assessment variable from  zero to ten where zero means ``not good at all in math" and ten ``good at math". \textit{Network} is a dummy taking the value of one if the person has family or friends who can support him/her for the next six months. \textit{Displacement}  has the value of one if the individual was internally displaced or zero otherwise. 
 \textit{House Damage} represents a self-assessment of the degree of damage of their house due to the earthquake, with five distinct categories of damage ranging from very low to total destruction of house. \textit{Risk Index} is composed by a self assessment risk and a stair case response following \citet{falk2018}. The higher the value the more risk living the individual is. 

\begin{table}[!htbp] \centering 
\resizebox{13cm}{!}{
\begin{tabular}{@{\extracolsep{5pt}}lcccc } 
\\[-1.8ex]\hline 
\hline \\[-1.8ex] 
    & \textbf{All Sample} & \multicolumn{2}{c}{\textbf{Non-Incentivized}} & \textbf{Incentivized}\\
\\[-0.9em] 
\cline{3-4} 
\\[-0.9em] 
    &       & \textbf{Control} & \textbf{Treatment} & (Experimental sample)\\ 
\cline{2-5}     
Female  \% & 0.483 & 0.455 & 0.503 & 0.503\\
\\[-1.8em]
Married \% & 0.636 & 0.573 & 0.680 & 0.669\\
\\[-1.8em]
Mean Age & 39.3 & 39.0 & 39.6 & 40.1\\
\\[-1.8em]
\textit{Education \%} &&&&\\
\\[-1.8em]
\quad Primary & 0.307 & 0.268 & 0.307 & 0.338\\
\\[-1.8em]
\quad Secondary & 0.352 & 0.358 & 0.352 & 0.368\\
\\[-1.8em]
\quad University & 0.322 & 0.358 & 0.322 & 0.278\\
\\[-1.8em]
\quad Graduate & 0.018 & 0.016 & 0.018 & 0.017\\
\\[-1.8em]
\textit{Income \%} &&&&\\
\\[-1.8em]
\quad $< 140 $ \euro  & 0.073 & 0.020 & 0.110 & 0.129\\
\\[-1.8em]
\quad 140 \euro - 339 \euro & 0.123 & 0.187 & 0.079 & 0.079\\
\\[-1.8em]
\quad 340 \euro - 669 \euro & 0.432 & 0.451 & 0.419 & 0.450\\
\\[-1.8em]
\quad 670 \euro - 999 \euro & 0.204 & 0.224 & 0.191 & 0.195\\
\\[-1.8em]
\quad $> 1000$ \euro & 0.168 & 0.118 & 0.202 & 0.146\\
\\[-1.8em]
Savings\% & 0.216 & 0.297 & 0.160 & 0.123\\
\\[-1.8em]
Math & 5.669 & 5.659 & 5.677 & 5.543\\
\\[-1.8em]
Network\% & 0.440 & 0.500 & 0.399 & 0.374\\
\\[-1.8em]
Displacement \% & 0.409 & 0.000 & 0.691 & 0.801\\
\\[-1.8em]
\textit{House Damage \%} &&&&\\
\quad Very low damage  & 0.281 & - & 0.281 & 0.166\\
\\[-1.8em]
\quad Low Damage & 0.197 & - & 0.197 & 0.232\\
\\[-1.8em]
\quad Medium Damage  & 0.171 & - & 0.171 & 0.192\\
\\[-1.8em]
\quad High Damage & 0.264 & - & 0.264 & 0.311\\
\\[-1.8em]
\quad Total Destruction & 0.084 & - & 0.084 & 0.099\\
\\[-1.8em]
Risk Index & 9.363 & 8.679 & 9.943 & 9.713\\
 \hline \\[-1.8ex] 
 Sample Size & 602 & 246 & 356 & 302\\
\\[-1.8em]
\hline 
\hline \\[-1.8ex] 
\end{tabular} 
}
  \caption[Descriptive Statistics for control and treatment groups]{Descriptive statistics. All sample that includes the control and treatment group. To test the first hypothesis the columns control and treatment were used. The last column corresponds to the sample used to analyse the second and  third hypothesis.} 
    \label{tab:summaryStatistics} 
\end{table}

\section{Results}
\label{section: results}

\subsection{Balance tests and matching procedure}
In this section we present the results pertaining  our first hypothesis (Section \ref{H1}), evaluating the effect of the earthquake on potential risk preferences via the GPS survey \citep{falk2018}. 
Table \ref{tab:summaryStatistics} shows the summary statistics  including the outcome variable,  \textit{risk index}. Table 2 presents all covariates for the control and treatment (earthquake) groups of the non-incentivized risk elicitation sample.  As we exploit a natural experiment thorough our experimental design, we we use the Propensity Score Matching (PSM) as an additional robustness check\footnote{We use the \textit{matchit} package using the software R for the matching procedure\citep{Ho2007}}. Specifically, we follow the optimal method which finds the matched samples with the smallest average absolute distance across all the matched pairs \citet{Ho2011}.

Table \ref{table:balanceTable} presents the balance before and after the matching procedure. The column \textit{F-Test} evaluates the significant difference between the control and treatment group before and after the matching\footnote{We also used the balance test described in \citet{Du2022}, the description is found in the Annex \ref{sec:balanceTableAnnex}}. We observe a variance in the average marriage variable between the treatment and control groups, which we deem not central to our research objectives, particularly considering recent experimental evidence regarding differences in risk preferences among married, unmarried couples, and individuals.\footnote{ \citet{BERNEDODELCARPIO2022101794} find that risk preferences of married couples and unrelated pairs are similar to the preferences of their constituent individuals.}.

Moreover, there is evidence that differences in marriage are pre-existing. Data from 2020 (before the earthquake), in the provinces of Adıyaman, Şanlıurfa, Hatay and Gazaintep, which are the origin of most of our affected population (and most severely affected by the earthquake), are amongst the ones with highest marriage rate when compared to the mean of the country\footnote{Marriage and Divorce Statistics for 2020, 25/02/2021 Statistical Press Report \url{https://data.tuik.gov.tr/Bulten/Index?p=Marriage-and-Divorce-Statistics-2020-37211&dil=2}. 
See Annex \ref{sec:MarriageRateComp} for details. }. 
After the matching process we detect no significant differences between groups, achieving balance across covariates. 
We have a sample of 492 individuals from both the treated and control groups, exhibiting similarity in the characteristics described in Table \ref{table:balanceTable}. We proceed to conduct all our analyses with both the matched and unmatched samples to ensure robustness.

\begin{table}[!ht] \centering 
\begin{adjustbox}{width=1\textwidth}
\begin{tabular}{@{\extracolsep{5pt}}p{1cm}cccccccccc} 
\\[-1.8ex]\hline 
\hline \\[-1.8ex] 
\\[-0.9em]
  & \multicolumn{5}{c}{\textbf{Balance table Before}} & \multicolumn{5}{c}{\textbf{Balance table After}}\\
\hline
\\[-1.8ex]
  & \multicolumn{2}{c}{\textbf{Control}} & \multicolumn{2}{c}{\textbf{Treatment}} & &\multicolumn{2}{c}{\textbf{Control}} & \multicolumn{2}{c}{\textbf{Treatment}} & \\
\\[-1.8ex]
\hline
\\[-1.8ex]
  & Mean & SD & Mean & SD & \textbf{F-test} & Mean & SD & Mean & SD & \textbf{F-test}\\
\\[-1.8ex]
Age & 39.03 & 13.93 & 39.65 & 13.68 & 0.29 & 39.03 & 13.93 & 39.87 & 14.68 & 0.42\\
\\[-1.8ex]
Female & 0.46 & 0.50 & 0.50 & 0.50 & 1.31 & 0.46 & 0.50 & 0.46 & 0.50 & 0.01\\
\\[-1.8ex]
Married & 0.57 & 0.50 & 0.68 & 0.47 & 7.20$^{***}$ & 0.57 & 0.50 & 0.58 & 0.49 & 0.07\\
\\[-1.8ex]
Education & 2.12 & 0.82 & 2.01 & 0.86 & 2.62 & 2.12 & 0.82 & 2.08 & 0.85 & 0.29\\
\\[-1.8ex]
Income & 3.35 & 1.31 & 3.23 & 1.70 & 0.96 & 3.35 & 1.31 & 3.33 & 1.76 & 0.02\\
\\[-1.8ex]
Math & 5.66 & 2.22 & 5.69 & 2.50 & 0.03 & 5.66 & 2.22 & 5.52 & 2.59 & 0.40\\
\hline \\[-1.8ex] 
N& 246 &  & 356 &  &  & 246 &  & 246&  &\\
\\[-1.8ex]\hline 
\hline \\[-1.8ex] 
\textit{Note:}  & \multicolumn{10}{r}{$^{*}$p$<$0.1; $^{**}$p$<$0.05; $^{***}$p$<$0.01} \\ 
\end{tabular} 
\end{adjustbox}
  \caption[Balance Table for the matching procedure]{Balance Table for the matching procedure. SD represents the Standard Deviation for each variable. The F-test show the significant differences between the control and group sample.} 
    \label{table:balanceTable} 
\end{table}

\subsection{Changes in Risk Preferences}

 We test the effect of the earthquake on risk preferences using our matched sample as per Hypotehsis 1. Regressions using the unmatched samples are presented in Annex \ref{sec:falkData2012}. We implement an Ordinary Least Square (OLS) regression with \textit{Risk Index} as a dependent variable (See Eq.\ref{eq:hipothesis1}).  Table \ref{tab: results1stHypothesis} shows the results of the regression. The first column shows that individuals who experienced the earthquake are significantly 1.4 points more risk-taking than those who did not. This is line with hypothesis 1, i.e, there is changes in risk preferences due to the impact of the earthquake. Moreover, at a baseline women are significantly less risk-taking than men. As per previous findings in the literature we proceed to test for potential heterogeneous effects of the earthquake across genders. For this we include an interaction between gender and our earthquake dummy that defines treatment (earthQ) in the second column. We find that overall significance of the \textit{earthQ} disappears and instead the interaction became highly significant. This implies that the effect of the earthquake on changes in risk preferences is driven by females.  Women who experience the earthquake are 3.887 points significantly more risk taking than those who did not (sum of $\beta_1$ and $\beta_3$ ). Second, women are less risk taking than men in the earthquake area\footnote{The average difference of risk taking among women and men from the earthquake area is derived from equation\ref{eq:hipothesis1} adding the coefficients $\beta_2$ and $\beta_3$ which results in a negative coefficient of -0.668.}. As a robustness check we replicate the same regression for the non-matched sample and our results  hold (See Annex \ref{sec:regBeforeMatching}).

\begin{equation}
\begin{split}
 Risk_i = \beta_0 + \beta_1 EarthQ_i + \beta_2 Female_i & + \beta_3 earthQ_i*Gender_i + \beta_2 Income_i + \\  
\beta_4 Married_i + \beta_5 Age_i + \beta_6 Education_i & + \beta_7 Saving_i + \beta_8 Math_i + \\ 
\beta_8 Network_i+ \epsilon_i       
\end{split}
\label{eq:hipothesis1}
\end{equation}


\begin{table}[!htbp] \centering 
\resizebox{12cm}{!}{
\begin{tabular}{@{\extracolsep{5pt}}lcc } 
\\
\hline 
\hline 
 & \multicolumn{2}{c}{\textit{Dependent variable:}} \\ 
\cline{2-3} 
				& \multicolumn{2}{c}{Risk Index} \\ [-0.9ex] 
					& \multicolumn{1}{c}{(1)} & \multicolumn{1}{c}{(2)}\\ 
\hline \\[-0.9ex] 
EarthQ & 1.448$^{**}$ & -0.577 \\[-0.9ex] 
  & (0.599) & (0.796) \\ [-0.9ex]
  Female & -2.922$^{***}$ & -5.132$^{***}$ \\[-0.9ex]   
  & (0.615) & (0.841) \\ [-0.9ex]
  EarthQ*Female &  & 4.464$^{***}$ \\ [-0.9ex]
  &  & (1.176) \\ [-0.9ex]
  Income & 0.797$^{***}$ & 0.778$^{***}$ \\[-0.9ex] 
  & (0.227) & (0.224) \\ [-0.9ex]
  Married & 0.095 & 0.291 \\ [-0.9ex]
  & (0.741) & (0.733) \\ [-0.9ex]
  Age & -0.041 & -0.052$^{*}$ \\ [-0.9ex]
  & (0.027) & (0.027) \\ [-0.9ex]
  Education & -0.151 & -0.211 \\ [-0.9ex]
  & (0.443) & (0.437) \\ [-0.9ex]
  Savings & 0.924 & 0.881 \\ [-0.9ex]
  & (0.805) & (0.794) \\ [-0.9ex]
  Math & 0.392$^{***}$ & 0.413$^{***}$ \\[-0.9ex] 
  & (0.139) & (0.137) \\ [-0.9ex]
  Network & -0.555 & -0.520 \\ [-0.9ex]
  & (0.625) & (0.617) \\ [-0.9ex]
  Constant & 6.949$^{***}$ & 8.342$^{***}$ \\[-0.9ex] 
  & (1.653) & (1.671) \\ [-0.9ex]
 \hline 
Observations & \multicolumn{1}{c}{486} & \multicolumn{1}{c}{486} \\ 
R$^{2}$ & \multicolumn{1}{c}{0.139} & \multicolumn{1}{c}{0.165} \\ 
Adjusted R$^{2}$ & \multicolumn{1}{c}{0.123} & \multicolumn{1}{c}{0.147} \\ 
Residual Std. Error & \multicolumn{1}{c}{6.491 (df = 476)} & \multicolumn{1}{c}{6.402 (df = 475)} \\ 
F Statistic & \multicolumn{1}{c}{8.572$^{***}$ (df = 9; 476)} & \multicolumn{1}{c}{9.374$^{***}$ (df = 10; 475)} \\ 
\hline  
\hline \\[-1.8ex] 
\textit{Note:}  & \multicolumn{2}{r}{$^{*}$p$<$0.1; $^{**}$p$<$0.05; $^{***}$p$<$0.01} \\ 
\end{tabular} 
}
  \caption[OLS regression with matched sample]{OLS regression with matched sample. The second column includes interaction among earthquake and gender.} 
  \label{tab: results1stHypothesis}
\end{table}

The significant differences in risk-taking among the earthquake and non-earthquake areas can potentially be attributed to systematic pre-existing differences between our treatment and control study areas. To assess pre-existing differences, we utilize data from \citet{falk2018}. We employ the same risk elicitation methodology as \citet{falk2018}, who gathered risk preferences in Turkey for the year 2012 in the same areas we used as control and treatment for our experimental design. We set up a regression using individual risk preferences collected in 2012 in Antalya as the control region. The treatment group is constructed using observations from Adana, Malatya, Hatay, and Gaziantep, gathering 136 observations and closely resembling the composition of our earthquake-treated sample in geographical distribution.

We conduct a OLS regression using the risk index as a dependent variable and available covariates such as gender, age, and math ability. With this we aim to check for the effect of pre-existing risk preferences across the regions or across gender among earthquake and non-earthquake areas in 2012. The results are presented in Table \ref{tab:robustCheck}, showing no differences in risk preferences across geographical location (earthQ) and gender. We conclude that there are no pre-existing differences between samples in 2012, supporting our main thesis that the existing differences in risk preferences can be attributed to the earthquake\footnote{Descriptive statistics about the data are found in Annex \ref{sec:falkData2012}}.

For Antalya, our control group, \citet{falk2018} have 40 observations. Hence, as a robustness check, we extended the control group to cover the cities of Samsun, Kastamonu, and Trabzon. These cities are comparable in characteristics to Antalya, in such way the sample size increased to a total of 136 observations. Specifically, these places have the same HDI as the regions affected by the earthquake in 2023, making the sample comparable. Overall, our previous results hold, finding no difference in risk preferences between our treatment group and the expanded control group on 2012 (See Annex Table \ref{tab:ols2012plus}).

\begin{table}[!htbp] 
\centering 
\resizebox{8cm}{!}{
\begin{tabular}{@{\extracolsep{5pt}}lc } 
\\[-1.8ex]\hline 
\hline  
 & \multicolumn{1}{c}{\textit{Dependent variable:}} \\ 
\cline{2-2} 
& \multicolumn{1}{c}{Risk Index} \\ 
\hline \\[-1.8ex] 
 EarthQ & -0.674 \\ 
  & (1.160) \\ 
  Gender & -1.190 \\ 
  & (0.948) \\ 
  Age & -0.104$^{***}$ \\ 
  & (0.035) \\ 
  Math & 0.345$^{*}$ \\ 
  & (0.178) \\ 
  Constant & 11.198$^{***}$ \\ 
  & (2.252) \\ 
 \hline \\[-1.8ex] 
Observations & \multicolumn{1}{c}{172} \\ 
R$^{2}$ & \multicolumn{1}{c}{0.103} \\ 
Adjusted R$^{2}$ & \multicolumn{1}{c}{0.081} \\ 
Residual Std. Error & \multicolumn{1}{c}{5.961 (df = 167)} \\ 
F Statistic & \multicolumn{1}{c}{4.786$^{***}$ (df = 4; 167)} \\ 
\hline 
\hline \\[-1.8ex] 
\textit{Note:}  & \multicolumn{1}{r}{$^{*}$p$<$0.1; $^{**}$p$<$0.05; $^{***}$p$<$0.01} \\ 
\end{tabular} 
}
  \caption[Differences in risk preferences between treatment and control]{OLS regression using the risk index as a dependent variable. Differences in risk preferences between individuals treatment and control area in 2012. The variable EarthQ is equal to one if individuals belongs to the earthquake area and zero otherwise. Gender is equal to one for women.} 
  \label{tab:robustCheck} 

\end{table}

\begin{figure}[hbt!]
  \centering
  \includegraphics[scale=0.55]{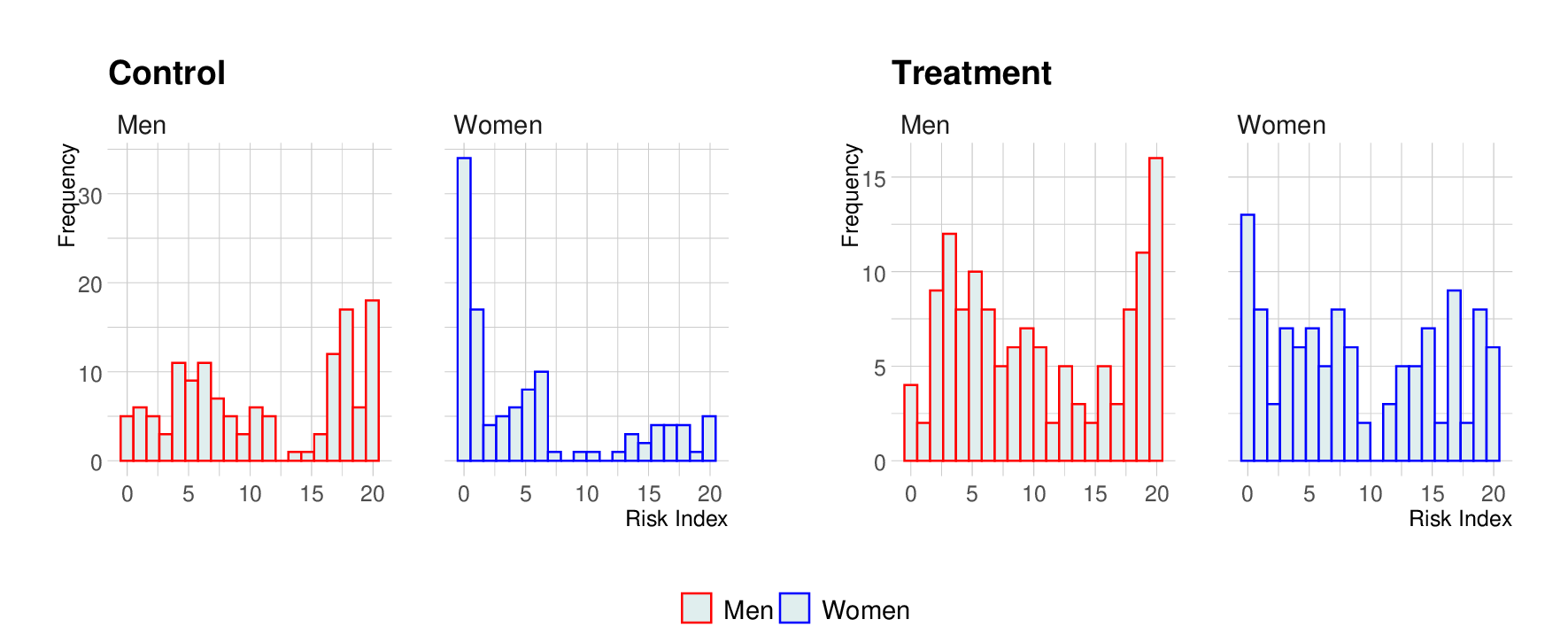}
  \caption[Histogram of the Risk Index variable for the matched sample]{Histogram of the Risk Index variable for the matched sample in the control and treatment regions by gender.}
  \label{fig:histogramGenderTreat}
\end{figure}

Figure \ref{fig:histogramGenderTreat} presents histograms of the risk index variable by gender for the control and treatment groups,. The figure illustrates that changes in risk preferences are driven by the targeted impact of the earthquake on women. The histograms reveal that women who experience the earthquake in the heavily affected region become more risk-taking than those in the control group. To gain a better understanding of potential drivers of this result, we perform further regressions with a sample of treated females (N=161) and males(N=160), using our Risk Index as the dependent variable. Table \ref{tab:genderDiffTreatControl} shows the results of the regression.

We explore several potential channels for our  results. Firstly, we include a variable \textit{Neg Income}, which has the value of one if there was a negative difference in income before and after the earthquake and zero otherwise. This variable aims to capture the effect of negative income shocks. Secondly, the variable \textit{Change Members} is a dummy with the value of one if the household experienced a change in the quantity of members of the family before and after the earthquake\footnote{Some households experience an increase in the quantity of members because people who loss members of their family can join another household.}. We add this variable to account for the loss of a family member, or conversely, new members who are now a responsibility for the household which might be driving changes in female risk preferences.

The within-treatment regression reveals that  a negative difference in income before and after the earthquake (income loss) is significantly correlated with more risk-taking for both men and women (See Table \ref{tab:genderDiffTreatControl}). This suggests that changes in risk preferences towards decreased loss aversion after the earthquake may stem from the generalized negative income shock. We find no statistical significance in the variable accounting for changes in the quantity of members of the family before and after the earthquake. 

\begin{table}[!htbp] \centering 
\resizebox{12cm}{!}{
\begin{tabular}{@{\extracolsep{5pt}}lcccc } 
\\[-1.8ex]\hline 
\hline \\[-1.8ex] 
 & \multicolumn{4}{c}{\textit{Dependent variable:}} \\ 
\cline{2-5} 
\\[-1.8ex] & \multicolumn{4}{c}{Risk Index} \\
 & \multicolumn{2}{c}{Female} & \multicolumn{2}{c}{Male} \\
& \multicolumn{1}{c}{(1)} & \multicolumn{1}{c}{(2)} & \multicolumn{1}{c}{(3)} & \multicolumn{1}{c}{(4)}\\ 
\hline \\[-1.8ex] 
 Age & -0.072 & -0.072 & -0.015 & -0.015 \\ 
  & (0.043) & (0.044) & (0.044) & (0.044) \\ 
  Education & -0.598 & -0.595 & -0.318 & -0.463 \\ 
  & (0.661) & (0.702) & (0.776) & (0.780) \\ 
  Married & 0.023 & 0.022 & -1.353 & -1.498 \\ 
  & (1.203) & (1.210) & (1.378) & (1.377) \\ 
  Neg, Income & 2.185$^{*}$ & 2.185$^{*}$ & 1.136 & 1.289 \\ 
  & (1.148) & (1.154) & (1.092) & (1.094) \\ 
  Property House & 0.165 & 0.164 & 0.854 & 0.937 \\ 
  & (1.087) & (1.091) & (1.069) & (1.067) \\ 
  Savings & 0.342 & 0.347 & 1.231 & 0.963 \\ 
  & (1.615) & (1.667) & (1.382) & (1.390) \\ 
  Trust & -1.735 & -1.733 & -2.422$^{**}$ & -2.235$^{*}$ \\ 
  & (1.082) & (1.102) & (1.158) & (1.161) \\ 
  Change Members & 0.205 & 0.204 & -3.082 & -3.107$^{*}$ \\ 
  & (2.341) & (2.350) & (1.875) & (1.869) \\ 
  Math & 0.618$^{***}$ & 0.617$^{***}$ & 0.314 & 0.229 \\ 
  & (0.221) & (0.226) & (0.218) & (0.226) \\ 
  Network &  & -0.015 &  & 1.612 \\ 
  &  & (1.164) &  & (1.138) \\ 
  Constant & 10.295$^{***}$ & 10.297$^{***}$ & 11.185$^{***}$ & 11.306$^{***}$ \\ 
  & (2.834) & (2.847) & (2.911) & (2.903) \\ 
 \hline \\[-1.8ex] 
Observations & \multicolumn{1}{c}{160} & \multicolumn{1}{c}{160} & \multicolumn{1}{c}{160} & \multicolumn{1}{c}{160} \\ 
R$^{2}$ & \multicolumn{1}{c}{0.132} & \multicolumn{1}{c}{0.132} & \multicolumn{1}{c}{0.088} & \multicolumn{1}{c}{0.100} \\ 
Adjusted R$^{2}$ & \multicolumn{1}{c}{0.080} & \multicolumn{1}{c}{0.074} & \multicolumn{1}{c}{0.033} & \multicolumn{1}{c}{0.039} \\ 
Residual Std. Error & \multicolumn{1}{c}{6.279 (df = 150)} & \multicolumn{1}{c}{6.300 (df = 149)} & \multicolumn{1}{c}{6.371 (df = 150)} & \multicolumn{1}{c}{6.349 (df = 149)} \\ 
F Statistic & \multicolumn{1}{c}{2.545$^{***}$ } & \multicolumn{1}{c}{2.275$^{**}$ } & \multicolumn{1}{c}{1.601 } & \multicolumn{1}{c}{1.651$^{*}$ } \\ 
            & \multicolumn{1}{c}{(df = 9; 150)} & \multicolumn{1}{c}{(df = 10; 149)} & \multicolumn{1}{c}{ (df = 9; 150)} & \multicolumn{1}{c}{ (df = 10; 149)} \\ 
\hline 
\hline \\[-1.8ex] 
\textit{Note:}  & \multicolumn{4}{r}{$^{*}$p$<$0.1; $^{**}$p$<$0.05; $^{***}$p$<$0.01} \\ 
\end{tabular}
} 
   \caption[OLS regression results for the sub-sample of earthquake area]{OLS regression results for the sub-sample of female and male of the earthquake area. } 
  \label{tab:genderDiffTreatControl} 

\end{table} 

\clearpage

\subsection{Risk Preferences and realized losses}


In this section, we present the results from our incentivized experiments \citep{Gneezy}, focusing exclusively on individuals heavily affected by the earthquake. The variable of interest is the percentage of money that individuals chose to invest in the risky option (\% invested in the bag. See \ref{fig:GeneezyPoters} ). This is a measure of risk-taking in this setting, as a higher proportion of investment indicates more risky behavior. We make use of a variable in our survey inquiring about the level of damage the house of the respondent perceived, measured from one to five, where five means total destruction. This variable is represented by \textit{House Damage} in our regressions. We consider that this variable encapsulates a significant aspect of \textit{realized losses} caused by the earthquake, at least constrained to the material domain. To test the effect of the magnitude of house damage on risk behavior, we also define medium damage and total damage as dummy variables, keeping the control group as the non-damage alternative. The medium damage encompasses the categories 2, 3, and 4 of the house damage variable, and total damage refers to the fifth category. With this set of variables and regressions, we aim to detect the relationship between realized losses, proxied by house damage, and risk preferences. We also aim to test if the loss magnitude has a differing effect among risk preferences in individuals affected by the earthquake. Annex \ref{sec:houseDamageAnnex} provides a detailed description of house damage in our sample.



Table \ref{tab: results1stHypothesis} shows the results of an OLS regression where the dependent variable is the percentage of endowment  that individuals invest in the risky choice. Regression (1)  shows that, on average , individuals who increase in one unit the house damage level invest less percentage of money in the risky option. When desegregating the house damage variable into categories proxying for the magnitude of the damage (Regression (2)), we find that people who experience a total destruction are significantly less risk taking than those who encounter no damage. Whereas, people who experienced medium damage exhibit less risk taking behavior than the baseline, however, this difference is not significant. This serves as tentative evidence that the magnitude of the realized loss may influence risk-taking behavior. 

\begin{table}[!htbp] \centering 

\resizebox{12cm}{!}{
\begin{tabular}{@{\extracolsep{5pt}}lcccc } 
\\[-1.8ex]\hline 
\hline \\[-1.8ex] 
 & \multicolumn{4}{c}{\textit{Dependent variable:}} \\ 
\cline{2-5} 
\\[-1.8ex] & \multicolumn{4}{c}{\% Investment in the Bag} \\ 
\\[-1.8ex] & \multicolumn{1}{c}{(1)} & \multicolumn{1}{c}{(2)} & \multicolumn{1}{c}{(3)} & \multicolumn{1}{c}{(4)}\\ 
\hline \\[-1.8ex] 
 House Damage & -0.036$^{**}$ &  & 0.003 &  \\ 
  & (0.018) &  & (0.023) &  \\ 
  Medium Damage &  & -0.080 &  &  \\ 
  &  & (0.062) &  &  \\ 
  Total Destruction &  & -0.182$^{**}$ &  &  \\ 
  &  & (0.090) &  &  \\ 
  Displacement &  &  & -0.205$^{***}$ & -0.198$^{***}$ \\ 
  &  &  & (0.077) & (0.058) \\ 
  Gender & -0.099$^{**}$ & -0.094$^{**}$ & -0.094$^{**}$ & -0.095$^{**}$ \\ 
  & (0.044) & (0.044) & (0.044) & (0.043) \\ 
  Age & -0.001 & -0.001 & -0.001 & -0.001 \\ 
  & (0.002) & (0.002) & (0.002) & (0.002) \\ 
  Married & -0.035 & -0.035 & -0.010 & -0.011 \\ 
  & (0.053) & (0.054) & (0.053) & (0.053) \\ 
  Education & 0.029 & 0.028 & 0.033 & 0.032 \\ 
  & (0.030) & (0.030) & (0.030) & (0.029) \\ 
  Income & 0.045$^{***}$ & 0.043$^{***}$ & 0.030$^{*}$ & 0.030$^{*}$ \\ 
  & (0.016) & (0.016) & (0.016) & (0.016) \\ 
  Employed & -0.076 & -0.075 & -0.067 & -0.067 \\ 
  & (0.054) & (0.054) & (0.053) & (0.053) \\ 
  Retiree & -0.098 & -0.100 & -0.113 & -0.113 \\ 
  & (0.079) & (0.079) & (0.078) & (0.078) \\ 
  Unemployed & -0.256$^{*}$ & -0.250 & -0.292$^{*}$ & -0.292$^{*}$ \\ 
  & (0.154) & (0.156) & (0.153) & (0.153) \\ 
  Constant & 0.685$^{***}$ & 0.665$^{***}$ & 0.758$^{***}$ & 0.762$^{***}$ \\ 
  & (0.123) & (0.124) & (0.124) & (0.121) \\ 
 \hline \\[-1.8ex] 
Observations & \multicolumn{1}{c}{298} & \multicolumn{1}{c}{298} & \multicolumn{1}{c}{298} & \multicolumn{1}{c}{298} \\ 
R$^{2}$ & \multicolumn{1}{c}{0.123} & \multicolumn{1}{c}{0.122} & \multicolumn{1}{c}{0.144} & \multicolumn{1}{c}{0.144} \\ 
Adjusted R$^{2}$ & \multicolumn{1}{c}{0.096} & \multicolumn{1}{c}{0.092} & \multicolumn{1}{c}{0.115} & \multicolumn{1}{c}{0.118} \\ 
Residual Std. Error & \multicolumn{1}{c}{0.367 (df = 288)} & \multicolumn{1}{c}{0.368 (df = 287)} & \multicolumn{1}{c}{0.363 (df = 287)} & \multicolumn{1}{c}{0.362 (df = 288)} \\ 
F Statistic & \multicolumn{1}{c}{4.489$^{***}$ } & \multicolumn{1}{c}{4.005$^{***}$ } & \multicolumn{1}{c}{4.841$^{***}$ } & \multicolumn{1}{c}{5.395$^{***}$ } \\ 
 & \multicolumn{1}{c}{ (df = 9; 288)} & \multicolumn{1}{c}{(df = 10; 287)} & \multicolumn{1}{c}{ (df = 10; 287)} & \multicolumn{1}{c}{(df = 9; 288)} \\ 
\hline 
\hline \\[-1.8ex] 
\textit{Note:}  & \multicolumn{4}{r}{$^{*}$p$<$0.1; $^{**}$p$<$0.05; $^{***}$p$<$0.01} \\ 
\end{tabular}
}
  \caption{OLS regression results realized losses on risk behavior} 
  \label{tab:results2ndHypothesis} 
 
\end{table} 

Realized losses, however, are not only restricted to the level of house damage or exclusive to the material domain, as per \citet{Imas2016}. In the context of an earthquake, realized losses comprise losing money but also any other medium of value. We argue that people who were internally displaced from the earthquake area most likely experienced a loss that goes beyond the strict definition of realized losses proxied by house damage (i.e., they might have physical assets which were a source of income such as an office, or machinery, or important sources of value that might go beyond the material realm). Hence, we include the variable \textit{Displacement} in regressions (3) and (4) in Table \ref{tab:results2ndHypothesis} to control for a potentially broader understanding of realized losses.

\begin{table}[!htbp] \centering 

\scalebox{0.7}{
\begin{tabular}{@{\extracolsep{5pt}}lcc} 
\\ 
\hline 
\hline \\
 & \multicolumn{2}{c}{\textit{Dependent variable:}} \\ 
\cline{2-3} 
\\ & \multicolumn{2}{c}{Displacement Behavior} \\ 
\\ & \multicolumn{1}{c}{(1)} & \multicolumn{1}{c}{(2)}\\ 
\hline  
  House Damage & 2.181$^{***}$ & 2.193$^{***}$ \\[-0.9ex] 
  & (0.392) & (0.396) \\ [-0.9ex]
  Gender & 0.156 & 0.185 \\ [-0.9ex]
  & (0.419) & (0.416) \\ [-0.9ex]
  Age & -0.001 & -0.002 \\ [-0.9ex]
  & (0.016) & (0.016) \\ [-0.9ex]
  Married & 0.384 & 0.446 \\ [-0.9ex]
  & (0.471) & (0.490) \\ [-0.9ex]
  Education & 0.577$^{*}$ & 0.588$^{*}$ \\ [-0.9ex]
  & (0.325) & (0.326) \\ [-0.9ex]
  Income & -0.864$^{***}$ & -0.861$^{***}$ \\ [-0.9ex]
  & (0.188) & (0.188) \\ [-0.9ex]
  Math & -0.051 & -0.047 \\ [-0.9ex]
  & (0.086) & (0.086) \\ [-0.9ex]
  Property & 0.068 & 0.108 \\ [-0.9ex]
  & (0.428) & (0.441) \\ [-0.9ex]
  No -Network & 0.835$^{**}$ & 0.848$^{**}$ \\[-0.9ex] 
  & (0.412) & (0.411) \\ [-0.9ex]
  Savings & -0.140 & -0.128 \\ [-0.9ex]
  & (0.521) & (0.524) \\ [-0.9ex]
  Change Members &  & 0.412 \\ [-0.9ex]
  &  & (0.911) \\ [-0.9ex]
  Constant & -1.526 & -1.656 \\[-0.9ex] 
  & (1.070) & (1.120) \\[-0.9ex]
 \hline \\[-0.9ex] 
Observations & \multicolumn{1}{c}{301} & \multicolumn{1}{c}{301} \\ 
Log Likelihood & \multicolumn{1}{c}{-30.880} & \multicolumn{1}{c}{-30.795} \\ 
Akaike Inf. Crit. & \multicolumn{1}{c}{83.761} & \multicolumn{1}{c}{85.590} \\ 
\hline 
\hline \\[-1.8ex] 
\textit{Note:}  & \multicolumn{2}{r}{$^{*}$p$<$0.1; $^{**}$p$<$0.05; $^{***}$p$<$0.01} \\ 
\end{tabular}
}
\caption[Drivers of displacement]{Drivers of displacement} 
  \label{tab:probitDisplacement} 
 
\end{table}

The displacement variable shows a significant and negative coefficient, meaning that people who were displaced are less risk-taking. This effect is highly significant and of higher magnitude than the house damage variable. This suggests that displacement captures a broader set of losses not exclusively defined by the individual's house damage. As a robustness check, we performed identical regressions in Table \ref{tab:results2ndHypothesis} using the risk index constructed from the GPS as dependent variable. The results still hold and are presented in Annex \ref{tab:realizedLossesRiskIndex}.




We further test the idea that the displacement variable potentially captures a broader understanding of realized losses. We conduct a probit regression on the displacement/ migration decision to understand its drivers (See Table \ref{tab:probitDisplacement}). We expect different aspects related to realized losses to be driving displacement for our previous statement to hold, i.e., displacement is driven by realized losses proxied by house damage and other lost of monetary mediums or values. Table \ref{tab:probitDisplacement} shows that the decision to internally migrate (or potentially being forced to leave, i.e., becoming internally displaced) is correlated with the level of house damage an individual experienced. Likewise, people who have no informal network (friends and family) to support them and have lower income are more likely to be displaced from their town of residence. Annex \ref{sec:damageMigration} presents the histogram of displacement per house damage level.

\subsection{High Order Risk Preferences}

\subsubsection{Prudence and Self protective behaviour}

Our third hypothesis focusses on treated individuals, i.e from the area heavily affected by the earthquake both displaced and not. We are interested in potential self-protective behaviors\footnote{(i.e., primary prevention, explained as behaviors/decisions reducing the likelihood of a loss occurring with the loss size being exogenous)}. We propose that internal displacement and/or internal migration can be understood as a form of self-protective behavior, and proceed to test this hypothesis using results from our incentivized experiments. 

\begin{figure}[ht!]
  \centering
  \includegraphics[scale=0.7]{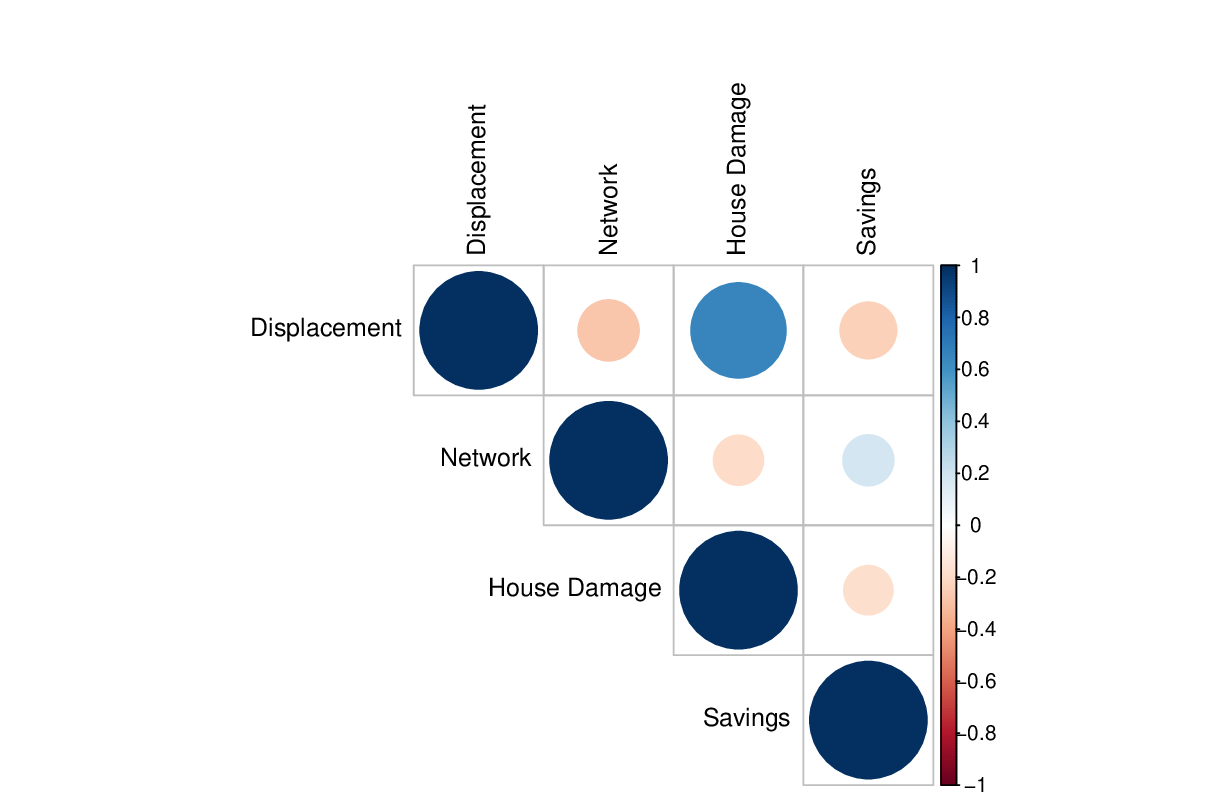}
  \caption{Summary correlations of covariates for prudence}
  \label{fig:corrPrudence}
\end{figure}

In the experiment, participants were presented with five choices, from which they could opt for either the prudent or imprudent option.  We generate a prudence variable, \textit{prudent choices}, which counts the number of prudent choices the individual makes. This variable is ordinal, ranging from one to five, with higher values indicating more prudent individuals. We employ an ordered probit regression.  Table \ref{tab:prudenceBehhavior} presents four regressions where the variables displacement, network and house damage are added sequentially. These three variables exhibit high correlation, as evidenced by previous results demonstrating the relationship between lack of network and house damage with displacement behavior (See Table \ref{tab:probitDisplacement} and figure \ref{fig:corrPrudence}). Additionally, stable income is a dummy variable with a value of one for individuals formally employed with a salary or retirees, and zero otherwise (where zero corresponds to self-employment). 

Results show a positive correlation between prudence and the decision to internally migrate/displacement, which is higher in both magnitude and significance than the house damage coefficient. This presents tentative evidence that the decision to internally migrate/displacement can be understood as a form of self protective behaviour, as it is most likely a way to ensure less exposure to current and future losses. We also observe little to no significance between prudence and variables related to displacement, albeit house damage exhibits a positive relatively weak association. Importantly, only age seems to be of importance when analysing prudence with individuals heavily affected by the earthquake.

\begin{table}[!htbp] \centering 
\scalebox{0.8}{
\begin{tabular}{@{\extracolsep{5pt}}lccc } 

\hline 
\hline 
 & \multicolumn{3}{c}{\textit{Dependent variable:}} \\ [-0.5ex]
\cline{2-4} 
 & \multicolumn{3}{c}{Prudent Choices} \\ [-0.5ex]
 & \multicolumn{1}{c}{(1)} & \multicolumn{1}{c}{(2)} & \multicolumn{1}{c}{(3)}\\ 
\hline \\[-1.8ex] 
 Age & 0.022$^{**}$ & 0.021$^{**}$ & 0.020$^{**}$ \\ 
  & (0.009) & (0.009) & (0.009) \\ 
  Married & 0.136 & 0.216 & 0.227 \\ 
  & (0.268) & (0.264) & (0.266) \\ 
  Education & 0.105 & 0.136 & 0.117 \\ 
  & (0.140) & (0.144) & (0.140) \\ 
  Displacement & 0.735$^{**}$ &  &  \\ 
  & (0.290) &  &  \\ 
  Network &  & -0.165 &  \\ 
  &  & (0.233) &  \\ 
  House Damage &  &  & 0.159$^{*}$ \\ 
  &  &  & (0.089) \\ 
  Gender & -0.140 & -0.175 & -0.131 \\ 
  & (0.218) & (0.220) & (0.220) \\ 
  Income Before & -0.003 & -0.057 & -0.046 \\ 
  & (0.083) & (0.081) & (0.081) \\ 
  Invest & 0.004 & 0.001 & 0.003 \\ 
  & (0.006) & (0.006) & (0.006) \\ 
  Savings & 0.248 & 0.156 & 0.204 \\ 
  & (0.326) & (0.329) & (0.326) \\ 
  Stable Income & -0.258 & -0.252 & -0.250 \\ 
  & (0.236) & (0.237) & (0.237) \\ 
 \hline \\[-1.8ex] 
Observations & \multicolumn{1}{c}{291} & \multicolumn{1}{c}{291} & \multicolumn{1}{c}{291} \\ 
\hline 
\hline \\[-1.8ex] 
\textit{Note:}  & \multicolumn{3}{r}{$^{*}$p$<$0.1; $^{**}$p$<$0.05; $^{***}$p$<$0.01} \\ 
\end{tabular} 
}
\caption{Prudence and Self-protective behaviour} 
\label{tab:prudenceBehhavior} 
\end{table}

\clearpage

\subsubsection{Prudence, Precautionary savings and self selection}

To analyse precautionary savings and prudence, we use the number of prudent choices individuals made in the experiment as a dependant variable . The higher values correspond to individuals which are more prudent. We analyze our results with  an ordered probit regression indicated in   table \ref{tab:prudenceBehhavior2}.

In the initial specification, presented in table \ref{tab:prudenceBehhavior},  we find no relation between prudence and savings. However, considering that prudence also plays a role in occupational choice, i.e., prudent individuals might prefer less risky income paths \citep{Fuchs-Schundeln2005}, we control for occupational self-selection using the variable stable income. We find that individuals who have stable incomes and savings are significantly more prudent. We repeat the regression across differing specifications with results being consistent. We also find significant negative association between prudence and having a stable income yet no savings. We believe our results indicate differentiated precautionary savings behavior in line with higher-order risk preferences, especially when accounting for self-selection into employment with riskier/less riskier income paths.

\begin{table}[!htbp] 
\centering 
\scalebox{0.8}{
\begin{tabular}{@{\extracolsep{5pt}}lcccc } 
\\[-1.8ex]\hline 
\hline \\[-1.8ex] 
 & \multicolumn{4}{c}{\textit{Dependent variable:}} \\ 
\cline{2-5} 
\\[-1.8ex] & \multicolumn{4}{c}{Prudence Behavior} \\ 
\\[-1.8ex] & \multicolumn{1}{c}{(1)} & \multicolumn{1}{c}{(2)} & \multicolumn{1}{c}{(3)} & \multicolumn{1}{c}{(4)}\\ 
\hline \\[-1.8ex] 
 Age & 0.022$^{**}$ & 0.023$^{**}$ & 0.022$^{**}$ & 0.021$^{**}$ \\ 
  & (0.009) & (0.009) & (0.009) & (0.009) \\ 
  Married & 0.136 & 0.096 & 0.180 & 0.189 \\ 
  & (0.268) & (0.270) & (0.266) & (0.268) \\ 
  Education & 0.105 & 0.129 & 0.162 & 0.140 \\ 
  & (0.140) & (0.141) & (0.145) & (0.141) \\ 
  Displacement & 0.735$^{**}$ & 0.717$^{**}$ &  &  \\ 
  & (0.290) & (0.291) &  &  \\ 
  Network &  &  & -0.159 &  \\ 
  &  &  & (0.233) &  \\ 
  House Damage &  &  &  & 0.140 \\ 
  &  &  &  & (0.089) \\ 
  Gender & -0.140 & -0.152 & -0.181 & -0.144 \\ 
  & (0.218) & (0.218) & (0.219) & (0.219) \\ 
  Income Before & -0.003 & -0.015 & -0.070 & -0.059 \\ 
  & (0.083) & (0.083) & (0.081) & (0.081) \\ 
  Invest & 0.004 & 0.003 & 0.0003 & 0.001 \\ 
  & (0.006) & (0.006) & (0.006) & (0.006) \\ 
  Savings & 0.248 & -0.316 & -0.425 & -0.339 \\ 
  & (0.326) & (0.429) & (0.430) & (0.432) \\ 
  Stable Income & -0.258 & -0.438$^{*}$ & -0.438$^{*}$ & -0.423$^{*}$ \\ 
  & (0.236) & (0.253) & (0.254) & (0.254) \\ 
  Savings:Stable Income &  & 1.265$^{**}$ & 1.328$^{**}$ & 1.213$^{*}$ \\ 
  &  & (0.632) & (0.638) & (0.637) \\ 
 \hline \\[-1.8ex]
Observations & \multicolumn{1}{c}{291} & \multicolumn{1}{c}{291} & \multicolumn{1}{c}{291} & \multicolumn{1}{c}{291} \\ 
\hline 
\hline \\[-1.8ex] 
\textit{Note:}  & \multicolumn{4}{r}{$^{*}$p$<$0.1; $^{**}$p$<$0.05; $^{***}$p$<$0.01} \\ 
\end{tabular}
} 
  \caption[Prudence and precautionary savings after the earthquake ]{Results of ordered probit regression regarding prudence and precautionary savings after the earthquake for affected individuals. The dependent variable is the number of prudent choices individuals made in the experiment. } 
      \label{tab:prudenceBehhavior2} 
\end{table}

\clearpage
\section{Discussion}
\label{section:discussion}

By investigating risk and higher preferences among individuals heavily affected by an earthquake, whether displaced or not, alongside a relevant control group, our aim is to contribute to the literature on changes in risk preferences following natural disasters. 
We find individuals who were heavily affected by the earthquake more risk taking in comparison with those not affected. This aligns with similar studies conducted in Japan following earthquakes \citep{Hanaoka2018}, Indonesia post-tsunamis \citep{Ingwersen2023}, and  the USA after hurricanes \cite{Eckel2009}. Importantly, these studies stand out for their methodological robustness, with both \citet{Hanaoka2018} and \citet{Ingwersen2023} having access to observations across time. 

We test for pre-existing differences using the approach outlined by \citet{falk2018} and data collected on the studied regions on 2012. We find no significant differences between our control and treatment groups (as defined by geographical location). This supports the notion that the differences in observed risk preferences are attributable to the earthquake. Yet, despite this finding which aligns with our proposition, the fact that the \citet{falk2018} data was collected in 2012, means that any potential impact between 2012 and 2023 could also be driving our results, which we consider as a limitation of this study. 

Our results withstand various robustness checks, including expanding the control group beyond 2012 to include other similar regions, and conducting regressions for both matched and unmatched samples. Importantly, when checking for heterogeneous effects of the earthquake across genders we found that our general finding, i.e. increased risk taking after the earthquake, is driven by females. Women who experienced the earthquake exhibit significantly more risk-taking behavior compared to those who did not. This finding aligns with the work of \citet{Eckel2009} and \citet{Abatayo2019} who find similar effects for females. 
We attempt to investigate potential drivers of women's behavior, finding that a negative change in income is associated with significantly increased risk-taking among all affected individuals, including men and women. Moreover, our design benefits from preference elicitation in a very short time frame after the natural disaster, approximately two months, as opposed to other similar research in the literature. For example, \citet{Abatayo2019},  elicited preferences 18 months after a typhoon, or \citet{Andrabi2017}, \citet{Ahsan2014} and \citet{Cassar2017} ran their experiments 3–4 years after a natural disaster struck. 

A great part of papers focusing on risk preferences and natural disasters finds that extreme events make individuals more risk-loving, while the remainder find the opposite effect\citep{Abatayo2019}. This existing contradiction in the literature on risk preferences after natural disasters is explained through  \citet{Imas2016} work, which is proposed as a reconciliation for the conflicting results. This is because \citet{Imas2016} show that realized and unrealized losses lead to opposite effects on risk preferences. Yet, \citet{Abatayo2019} 
 suggested that most effects of natural catastrophes are realized, thus, one would expect consistent results towards increased risk aversion. Moreover, \citet{Imas2016}'s results might not be applicable to developing contexts, or perhaps not to fieldwork, as \citet{Imas2016} primarily operates in controlled settings with stock traders. 
 
We test if realized losses, in the form of house damage after the earthquake, indeed leads to risk aversion. Interestingly, we consistently found that realized losses lead to risk aversion, in line with \citet{Imas2016}. Furthermore, this effect appears to be particularly pronounced among individuals who experienced a total loss of their home, suggesting that the effect intensifies with the magnitude of total loss. This result in itself highlights the existing contradiction in the literature. Overall, we find that people who experienced the earthquake became more risk-taking than people who did not, however, heavily affected individuals who suffered catastrophic losses became more risk-averse. 

We further examine the role of losses by investigating the decision for internal migration following the earthquake. We believe that internal migration encompasses losses that extend beyond the exclusive realm of material losses related to one's house. The decision to migrate (or being displaced) is likely correlated with other realized losses, which could also be non-monetary. We consistently found that internally displaced individuals tend to be more risk-averse than those who experienced the earthquake but did not migrate, further reinforcing this point.

To ensure the robustness of our assumption regarding internal migration and realized losses, we inquire into the drivers of internal displacement. We discover that a lack of informal networks, lesser income, and the level of house damage is correlated to internal migration. This suggests that internal displacement could serve as a suitable  proxy for a broader understanding of realized losses in our setting.  According to the Internation Organization for Migration (IOM), people who move voluntarily within a country can be regarded as internal migrants, including several reasons formally and informally. If their movement is forced, individuals are referred to as Internally Displaced Persons (IDP). We consider that a significant part of affected individuals in our sample are IDP, they were forced to move due to conditions, whilst others are most likely internal migrants, as their decision might be a choice.

To our knowledge, we present the first examination of higher-order risk preferences in the context of material losses and internal displacement following a natural catastrophe. This contributes to the very limited empirical literature on prudence \citet{Lugilde2019}. We initially explored the concept of self-protective behavior and prudence, positing that the decision to internally migrate can be understood as a self-protective measure. I.e, leaving an earthquake-prone area likely reduces the likelihood of experiencing a future loss. We found that prudence is positively correlated with the decision to internally migrate, as stated above, this internal migration may be a forceful manner for a subgroup of our sample. 

\citet{Eeckhoudt2005} demonstrated that a prudent individual expends less preventive effort than a risk-neutral one.  \citet{Courbagea2006} extended this model into the health context, where prudence also leads to lower optimal prevention efforts. However, \citet{Menegatti2009} expands the original model from one to two periods, arguing that in many prevention situations, the preventive effort precedes its effect on the probability. In contrast to the results of one-period models, he found that prudence leads to more prevention. Our results show a positive correlation between prudence and migration/internal displacement, which we believe can be understood as a form of self-protective behavior. This is in line with \citet{Menegatti2009}'s theoretical results. However, \citet{Peter2017} showed that once saving decisions are incorporated (i.e., they are endogenous), results of the one-period models are restored. In other words, prudence should lead to less prevention in the sense of self-protection. Yet, for this to hold, individuals must be able to optimize both their savings (and, thus, also consumption) and prevention decisions. We believe this is not the case for individuals facing constraints imposed by a natural catastrophe, as our results align with \citet{Menegatti2009} theoretical findings.

Moreover, we explored the potential role of precautionary savings  and higher order risk preferences in our study. Prudent individuals, expecting future income shocks, typically accumulate precautionary savings to smooth consumption. However, in our initial analysis, we found no clear correlation between savings and prudence. This seems counter-intuitive, as, in the absence of complete insurance, one would expect that expected future income shocks drive prudent individuals to build up precautionary savings to avoid wide fluctuations in consumption.

Nevertheless, it is essential to consider that risk aversion not only influences savings behavior but also plays a role in occupational choice. Prudent individuals might opt for occupations associated with less risky income paths, while less prudent may prefer occupations with higher income risk \citep{Fuchs-Schundeln2005}. In line with these findings on self-selection and precautionary savings \citep{Fuchs-Schundeln2005}, we examine potential self-selection of prudent individuals into occupations linked with less risky income paths. Our analysis provided suggestive evidence of self-selection, indicating that individuals (heavily affected by the earthquake) who self-select into stable income jobs and have savings are significantly more prudent than those who are self-employed and lack savings. While our results regarding higher-order risk preferences in catastrophe settings cannot be regarded as causal findings, we believe to present novel suggestive evidence in the field of empirical research on prudence related behaviors \citep{Lugilde2019}, such as primary prevention and precautionary savings in the context of a natural disaster.

Importantly, there is evidence that conclusions derived from changes in risk preference from incentivized experiments have a relation with `real-life' risk-taking behaviors. \citet{Cameron2015} find that individuals in villages in Indonesia that suffered a flood or earthquake within the past three years display higher levels of risk aversion in comparison to individuals unaffected by such events. They examine the extent to which behavior in the risk experiments is correlated with such `real-life' risk-taking as opening a new business or changing jobs. They provide evidence that more risk-averse individuals are less likely to take these types of risky decisions. This could potentially imply long run economic impacts that could be attributable to changes of risk preferences from natural disasters. 
 
Timing is an important aspect to consider. A notable limitation of our study is the absence of continuous observations across time for the same individuals. As a result, we are unable to provide insights into the long-term effects of catastrophes on risk preferences or the permanence of our findings. In this aspect, the literature finds conflicting evidence on the permanence of changes in risk preferences across time due to natural disasters. \citet{Hanaoka2018} finds persistent changes in risk preferences after an earthquake in  Japan, whereas \citet{Ingwersen2023}  in a tsunami context finds that changes in risk preferences are short-lived: starting a year later. To our knowledge there is no evidence of differences in willingness to take risk between the two groups. We believe that the permanence of changes in risk preferences after a natural disaster, and their correlation to economic behaviour are important avenues for future research.

\section*{Acknowledgements}
 This paper benefited greatly from discussions with Grischa Perino, Andreas Lange, Moritz Drupp, Robbert Jan Schaap and Marta Talevi.  We thank the Environmental and Resource Economics seminar at the University of Hamburg and the University of Kiel for helpful suggestions. We specifically thank Ümit Yakar for additional help on language, translations and logistics to perform our research in the field. Moreover, we thank our team of local researchers for the support in the implementation of the research. Finally we thank the team at the Briq institute, specifically Prof. Armin Falk and Dr. Markus Antony for the openness to share data for this research.

\section*{Funding}
\noindent We are grateful for the funding received from the  Deutsche Forschungsgemeinschaft (DFG, German Research Foundation) under Germany‘s Excellence Strategy – EXC 2037 ’CLICCS - Climate, Climatic Change, and Society’ – Project Number: 390683824, contribution to the Center for Earth System Research and Sustainability (CEN) of Universität Hamburg. Likewise we are grateful for the financial support provided by the Graduate School of the  School of Economics and Social Sciences.

\renewcommand{\thesubsection}{A-\arabic{subsection}}
\setcounter{subsection}{0}
\renewcommand{\thefigure}{A-\arabic{figure}}
\setcounter{figure}{0}
\renewcommand{\thetable}{A-\arabic{table}}
\setcounter{table}{0}

\newpage
\begin{subappendices}


\textbf{APPENDIX}

\section{Risk assessment}
\label{sec:riskTreeFalk}
\begin{figure}[ht!]
  \centering
  \includegraphics[scale=0.9]{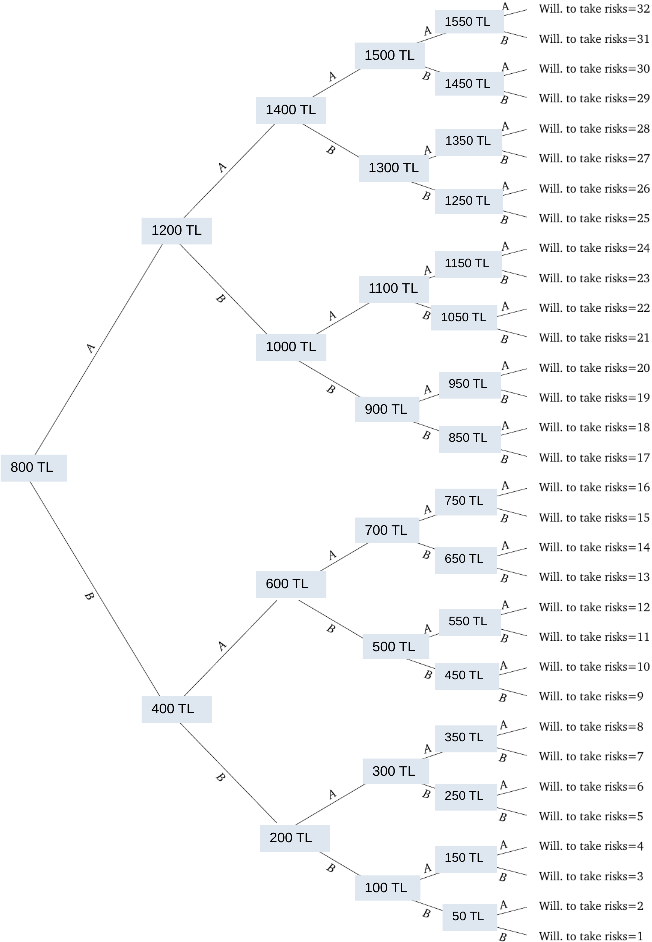}
  \caption[Risk tree to assess risk behavior]{Risk Tree used in the experiment to assess risk behavior. The staircase risk task involved a decision tree, where numbers represented guaranteed payments, "A" denoted the choice of a lottery, and "B" indicated the choice of a certain payment. The procedure operated as follows: participants were initially asked whether they preferred a guaranteed payment of 800 lira or a 50:50 chance of either receiving 1500 lira or nothing. If they chose the safe option ("B"), the subsequent guaranteed amount inquired decreased to 400 lira. Conversely, if they opted for the gamble ("A"), the assured amount was raised to 1200 lira. This pattern continued throughout the tree, following the same rationale. The risk tree values were based on those indicated in \citep{falk2018} taken in 2012. These values were converted in real values for the time of the experiment in 2023.}
  \label{fig:riskTree}
\end{figure}

\newpage
\begin{figure}[ht!]
  \centering
  \includegraphics[scale=1.1]{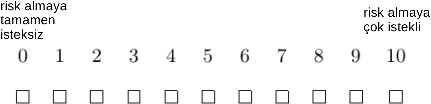}
  \caption[Self assessment question about the willingness to take risk]{Self assessment question about the willingness to take risk. Participants were inquired: Please tell me, in general, how willing or unwilling you are to take risks. Please use a scale from 0 to 10, where 0 means “completely unwilling to take risks” and a 10 means you are “very willing to take risks”.}
  \label{fig:riskWilligness}
\end{figure}

 \section{Risk experiment based on Gneezy and Potters}
 \label{sec:GenezzeyPotterImage}
\begin{figure}[ht!]
  \centering
  \includegraphics[scale=1.3]{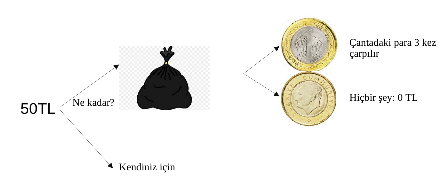}
  \caption[Risk experiment]{Structure of the risk experiment based on Gneezy and Potters}
   \label{fig:GeneezyPoters}
\end{figure}

 \section{Prudence experiment}
 \label{sec:EeckhoudtSchlesinger}
\begin{figure}[ht]
  \centering
  \includegraphics[scale=0.6]{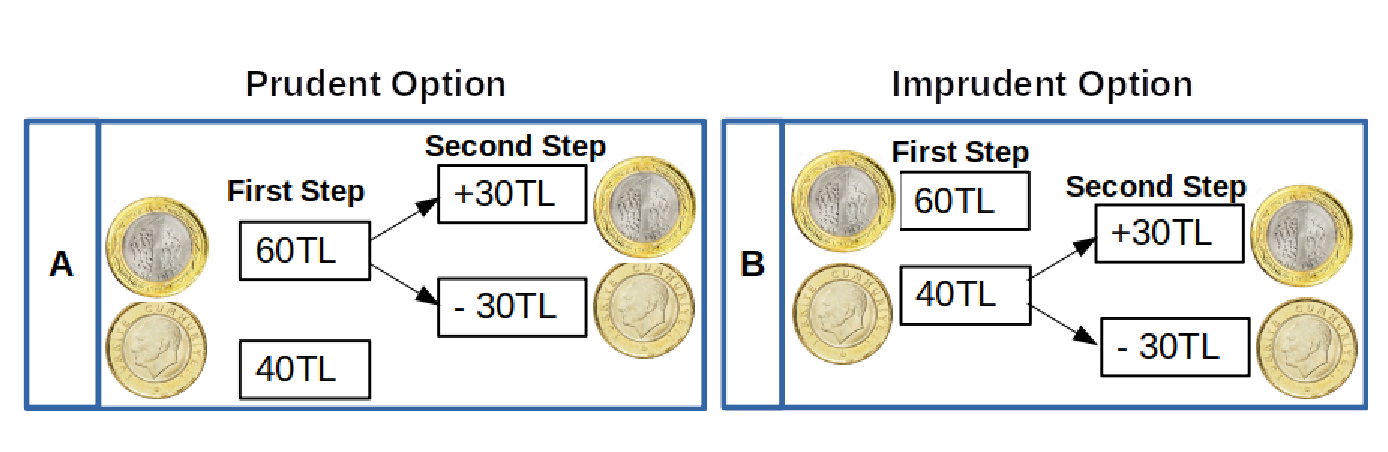}
  \caption[Prudence experiment]{Structure of the prudence experiment based on \citep{Schaap2021}. The figure shows the two options in the base line of the choice task.}
  \label{fig:EeckhoudtSchlesinger}
\end{figure}

\newpage

 \section{Balance Table}
 \label{sec:balanceTableAnnex}

\begin{table}[!htbp] \centering 
\begin{adjustbox}{width=0.8\textwidth}
\begin{tabular}{@{\extracolsep{5pt}}p{1.5cm}cccccccccc} 
\hline 
\hline  
\\[-0.9em]
  & \multicolumn{5}{c}{\textbf{Balance table Before}} & \multicolumn{5}{c}{\textbf{Balance table After}}\\
\hline
  & \multicolumn{2}{c}{\textbf{Control}} & \multicolumn{2}{c}{\textbf{Treatment}} & &\multicolumn{2}{c}{\textbf{Control}} & \multicolumn{2}{c}{\textbf{Treatment}} & \\
\hline
\\[-1.8ex]
  & Mean & SD & Mean & SD & \textbf{Diff} & Mean & SD & Mean & SD & \textbf{Diff}\\[-0.9ex]
Age & 39.03 & 13.93 & 39.65 & 13.68 & 0.04 & 39.03 & 13.93 & 39.87 & 14.68 & 0.06\\[-0.9ex]
Gender & 0.46 & 0.50 & 0.50 & 0.50 & 0.10 & 0.46 & 0.50 & 0.46 & 0.50 & 0.01\\[-0.9ex]
Married & 0.57 & 0.50 & 0.68 & 0.47 & 0.22 & 0.57 & 0.50 & 0.58 & 0.49 & 0.03\\[-0.9ex]
Education & 2.12 & 0.82 & 2.01 & 0.86 & 0.14 & 2.12 & 0.82 & 2.08 & 0.85 & 0.05\\[-0.9ex]
Income & 3.35 & 1.31 & 3.23 & 1.70 & 0.08 & 3.35 & 1.31 & 3.33 & 1.76 & 0.01\\[-0.9ex]
Math & 5.66 & 2.22 & 5.69 & 2.50 & 0.01 & 5.66 & 2.22 & 5.52 & 2.59 & 0.06\\[-0.9ex]
\hline  
N& 246 &  & 356 &  &  & 246 &  & 246&  &\\
\hline 
\hline 
\textit{Note:}  & \multicolumn{10}{r}{$^{*}$p$<$0.1; $^{**}$p$<$0.05; $^{***}$p$<$0.01} \\ 
\end{tabular} 
\end{adjustbox}
  \caption[Balance Table for the matching procedure]{Balance Table for the matching procedure. To test for imbalance we used the software R, specifically the package 'stddiff' . Imbalance is usually defined as a \textit{Diff} greater than 0.1 or 0.2} 
    \label{table:balanceTableAnnex} 
\end{table}

 \section{Average marriage rate}
 \label{sec:MarriageRateComp}

\begin{table}[!htbp] \centering
\centering
\scalebox{0.8}{
\begin{tabular}{lcc}
\hline
Year & Antalya & Mean Treatment\\
\hline
2012 & 8.28 & 10.09\\[-0.9ex]
2013 & 8.00 & 10.12\\[-0.9ex]
2014 & 7.91 & 9.98\\[-0.9ex]
2015 & 7.97 & 9.30\\[-0.9ex]
2016 & 7.61 & 8.74\\[-0.9ex]
2017 & 7.02 & 8.29\\[-0.9ex]
2018 & 6.76 & 8.31\\[-0.9ex]
2019 & 6.76 & 7.87\\[-0.9ex]
2020 & 6.17 & 7.16\\[-0.9ex]
2021 & 6.59 & 8.02\\[-0.9ex]
2022 & 6.80 & 7.39\\[-0.9ex]
\hline
t- test & \multicolumn{2}{r}{t = -3.611, df = 17.36$^{***}$} \\
\hline\\
\textit{Note:}  & \multicolumn{2}{r}{$^{*}$p$<$0.1; $^{**}$p$<$0.05; $^{***}$p$<$0.01} \\ 
\end{tabular}
}
\caption[Average marriage rate per 1000 population among 2012 and 2022]{Average marriage rate per 1000 population among 2012 and 2022 for Antalya (control) and the treatment provinces including Adiyaman, Gaziantep, Hatay, Malatya, Kahramanmaraş, Şanlıurfa and Osmaniye. The t-test in the last row shows that there are significant differences among these two regions. Source: Turkish Statistical Institute (\textcolor{red}{XXXXX})
}
\end{table}

\clearpage

\section{Regression with the whole sample - Before matching}
\label{sec:regBeforeMatching}

\begin{table}[!htbp] \centering 
 \scalebox{0.7}{
\begin{tabular}{@{\extracolsep{5pt}}lcc } 
\\[-1.8ex]\hline 
\hline 
 & \multicolumn{2}{c}{\textit{Dependent variable:}} \\ 
\cline{2-3} 
 & \multicolumn{2}{c}{Risk Index} \\ 
 & \multicolumn{1}{c}{(1)} & \multicolumn{1}{c}{(2)}\\ 
\hline \\[-1.8ex] 
 EarthQ & 1.531$^{***}$ & -0.697 \\ 
  & (0.550) & (0.739) \\ 
  Gender & -2.407$^{***}$ & -5.202$^{***}$ \\ 
  & (0.544) & (0.828) \\ 
    EarthQ*Gender &  & 4.706$^{***}$ \\ 
  &  & (1.063) \\ 
  Income & 0.703$^{***}$ & 0.715$^{***}$ \\ 
  & (0.203) & (0.200) \\ 
  Married & 0.344 & 0.364 \\ 
  & (0.663) & (0.653) \\ 
  Age & -0.034 & -0.043$^{*}$ \\ 
  & (0.024) & (0.023) \\ 
  Education & 0.136 & 0.077 \\ 
  & (0.395) & (0.389) \\ 
  Savings & 0.800 & 0.752 \\ 
  & (0.736) & (0.724) \\ 
  Math & 0.352$^{***}$ & 0.347$^{***}$ \\ 
  & (0.122) & (0.120) \\ 
  Network & -0.503 & -0.445 \\ 
  & (0.565) & (0.556) \\ 
  Constant & 6.275$^{***}$ & 7.980$^{***}$ \\ 
  & (1.468) & (1.496) \\ 
 \hline \\[-1.8ex] 
Observations & \multicolumn{1}{c}{594} & \multicolumn{1}{c}{594} \\ 
R$^{2}$ & \multicolumn{1}{c}{0.115} & \multicolumn{1}{c}{0.144} \\ 
Adjusted R$^{2}$ & \multicolumn{1}{c}{0.101} & \multicolumn{1}{c}{0.129} \\ 
Residual Std. Error & \multicolumn{1}{c}{6.424 (df = 584)} & \multicolumn{1}{c}{6.324 (df = 583)} \\ 
F Statistic & \multicolumn{1}{c}{8.443$^{***}$ (df = 9; 584)} & \multicolumn{1}{c}{9.802$^{***}$ (df = 10; 583)} \\ 
\hline  
\hline \\[-1.8ex] 
\textit{Note:}  & \multicolumn{2}{r}{$^{*}$p$<$0.1; $^{**}$p$<$0.05; $^{***}$p$<$0.01} \\ 
\end{tabular} 
}
 \caption[OLS regression with the whole sample before matching]{OLS regression with the whole sample before matching} 
\end{table} 

\clearpage
\section{Difference among earthQ and non-earthQ area in 2012}
\label{sec:falkData2012}

\begin{table}[h!]
\centering
\scalebox{0.9}{
\begin{tabular}{lcccccccc}
\hline
  & \multicolumn{2}{c}{\textbf{Control}} & \multicolumn{2}{c}{\textbf{Control Plus}} & \multicolumn{2}{c}{\textbf{Treatment}} & &\\
\hline
\\
  & Mean & SD & Mean & SD & Mean & SD & F-Test-A & F-Test-B\\[-0.9ex]
Gender & 0.42 & 0.50 & 0.57 & 0.50 & 0.61 & 0.49 & 4.38$^{**}$ & 0.38\\[-0.9ex]
Age & 45.20 & 14.69 & 39.55 & 13.72 & 35.88 & 13.58 & 14.04$^{***}$ & 4.93$^{**}$\\[-0.9ex]
Self Risk & 4.75 & 2.71 & 5.34 & 2.69 & 5.14 & 2.96 & 0.56 & 0.32\\[-0.9ex]
Stair Risk & 10.14 & 10.66 & 11.04 & 11.08 & 10.95 & 12.36 & 0.14 & 0.00\\[-0.9ex]
Math & 4.15 & 2.73 & 4.78 & 2.81 & 5.41 & 2.69 & 6.55 & 3.58\\[-0.9ex]
Risk Index & 7.30 & 5.35 & 8.03 & 5.66 & 7.89 & 6.43 & 0.28 & 0.04\\[-0.9ex]
N & 40 &  & 136 & &136 &  && \\[-0.9ex]
\hline
\textit{Note:}  & \multicolumn{5}{r}{$^{*}$p$<$0.1; $^{**}$p$<$0.05; $^{***}$p$<$0.01} \\ 
\end{tabular}
}
 \caption[ Descriptive statistics for control and treatment groups during4 2012]{ Descriptive statistics for groups in the control and treatment regions in the year 2012 based on data from \cite{falk2018}. The control columns refer to Antalya, but as we have only 40 observations we added to the control group the places of Kastamonu, Trabzon and Samsun which also have the same HDI. The group with the added cities corresponds to the control plus columns. The treatment regions gather Adana, Malatya, Hatay and Gaziantep, which are the regions affected by the earthquake in 2023.
 The F-Test-A shows the significant differences among control and treatment regions. The F-Test-B shows the differences among the control plus and the treatment regions.   }
\end{table}

\begin{table}[!htbp] \centering
 \scalebox{0.7}{
\begin{tabular}{@{\extracolsep{5pt}}lc} 
\hline
\hline
  & \multicolumn{1}{c}{\textit{Dependent variable:}} \\ 
\cline{2-2} 
\\[-1.8ex] & \multicolumn{1}{c}{Risk Index} \\ 
\hline \\[-1.8ex] 
 EarthQ & -0.597 \\ 
  & (0.727) \\ 
  Gender & -1.153 \\ 
  & (0.744) \\ 
  Age & -0.099$^{***}$ \\ 
  & (0.028) \\ 
  Math & 0.217 \\ 
  & (0.138) \\ 
  Constant & 11.621$^{***}$ \\ 
  & (1.650) \\ 
 \hline \\[-1.8ex] 
Observations & \multicolumn{1}{c}{268} \\ 
R$^{2}$ & \multicolumn{1}{c}{0.079} \\ 
Adjusted R$^{2}$ & \multicolumn{1}{c}{0.065} \\ 
Residual Std. Error & \multicolumn{1}{c}{5.863 (df = 263)} \\ 
F Statistic & \multicolumn{1}{c}{5.652$^{***}$ (df = 4; 263)} \\ 
\hline 
\hline \\[-1.8ex] 
\textit{Note:}  & \multicolumn{1}{r}{$^{*}$p$<$0.1; $^{**}$p$<$0.05; $^{***}$p$<$0.01} \\ 
\end{tabular} 
}
  \caption[OLS regression with the extended sample]{OLS regression with the extended sample: The control plus and the treatment regions.} 
  \label{tab:ols2012plus} 

\end{table}

\clearpage
\section{Histograms per level of house damage}
\label{sec:houseDamageAnnex}

\begin{figure}[ht]
  \centering
  \includegraphics[scale=0.8]{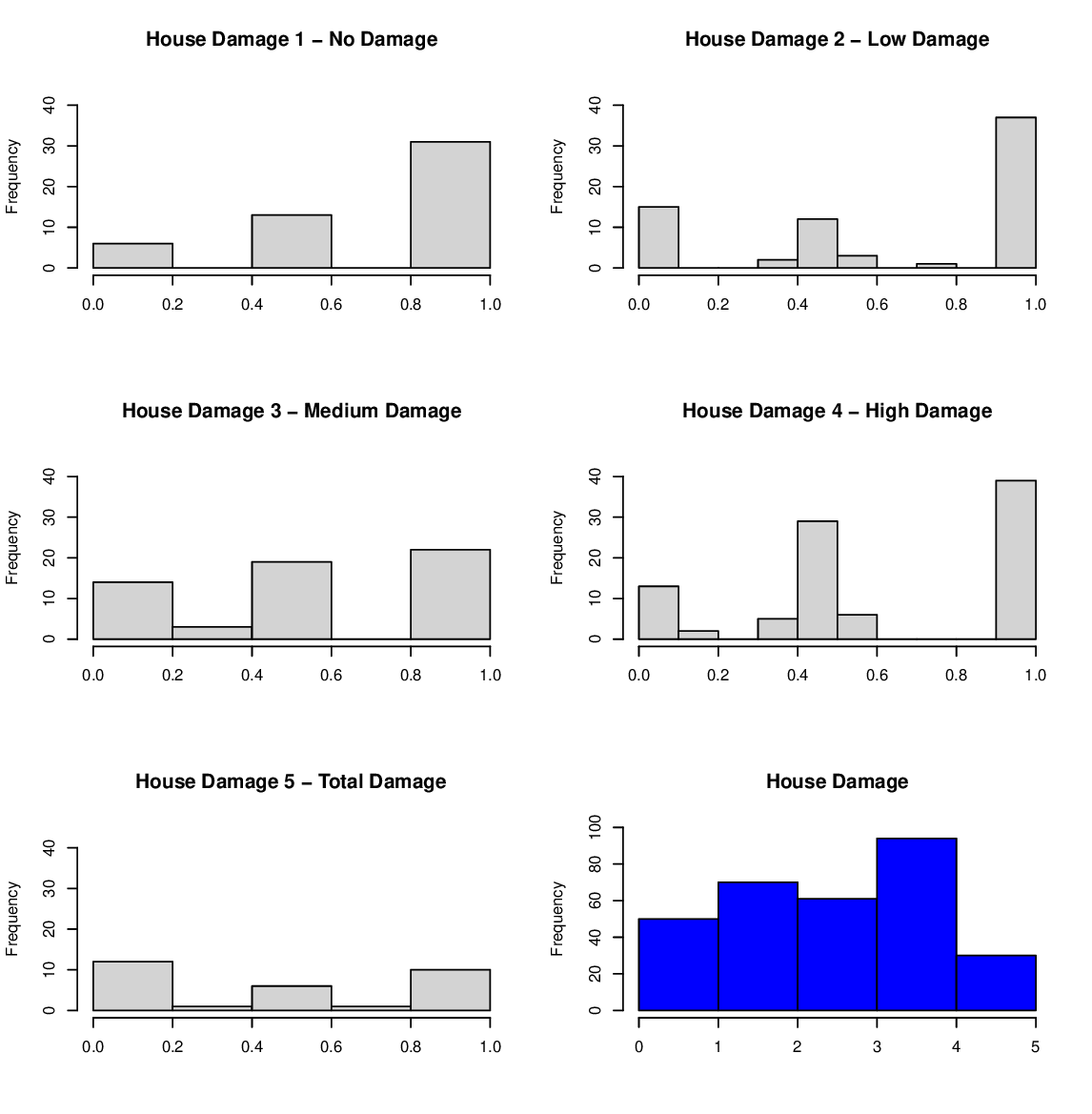}
  \caption[Histograms of proportion of initial endowment]{Histograms of proportion of initial endowment invested in the bag per level of damage. X- axis: for proportion invested in the Bag. The blue histogram shows the distribution of house damage in the whole sample.}  
\end{figure}

\clearpage

\section{Histogram displacement per level of damage}
\label{sec:damageMigration}

\begin{figure}[ht]
  \centering
  \includegraphics[scale=0.8]{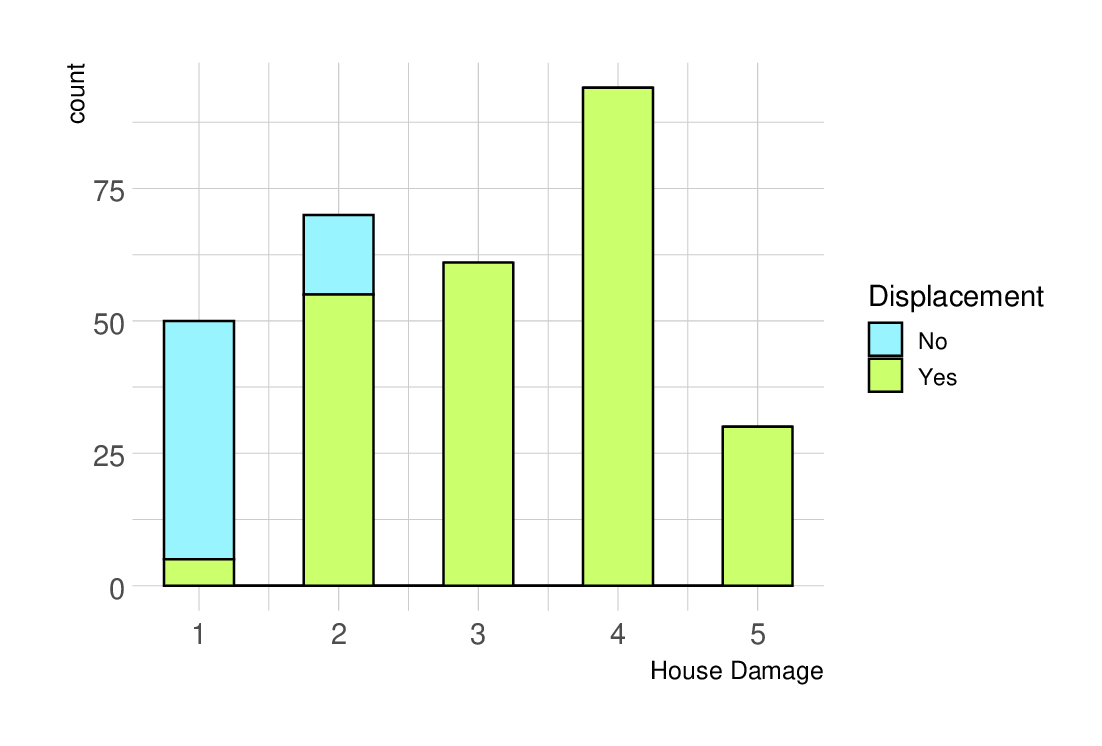}
  \caption[Histogram displacement per level of damage]{Histogram displacement per level of damage: 5: total damage. 1: no damage.}
\end{figure}

\clearpage

\section{Realized losses on risk behavior}
\label{sec:realizedLossesRiskIndex }

\begin{table}[!htbp] \centering 
\scalebox{0.7}{
\begin{tabular}{@{\extracolsep{5pt}}lcccc } 
\\[-1.8ex]\hline 
\hline \\[-1.8ex] 
 & \multicolumn{4}{c}{\textit{Dependent variable:}} \\ 
\cline{2-5} 
\\[-1.8ex] & \multicolumn{4}{c}{Risk Taking} \\ 
\\[-1.8ex] & \multicolumn{1}{c}{(1)} & \multicolumn{1}{c}{(2)} & \multicolumn{1}{c}{(3)} & \multicolumn{1}{c}{(4)}\\ 
\hline \\[-1.8ex] 
 House Damage & -0.727$^{**}$ &  & -0.391 &  \\ 
  & (0.310) &  & (0.405) &  \\ 
  Medium Damage &  & -1.795 &  &  \\ 
  &  & (1.090) &  &  \\ 
  Total Destruction &  & -3.218$^{**}$ &  &  \\ 
  &  & (1.595) &  &  \\ 
  Displacement &  &  & -1.755 & -2.604$^{**}$ \\ 
  &  &  & (1.363) & (1.043) \\ 
  Gender & -0.976 & -0.892 & -0.936 & -0.903 \\ 
  & (0.772) & (0.775) & (0.772) & (0.771) \\ 
  Age & -0.069$^{*}$ & -0.072$^{*}$ & -0.069$^{*}$ & -0.073$^{*}$ \\ 
  & (0.037) & (0.038) & (0.037) & (0.037) \\ 
  Married & -0.683 & -0.615 & -0.472 & -0.336 \\ 
  & (0.931) & (0.944) & (0.944) & (0.933) \\ 
  Education & -0.280 & -0.271 & -0.240 & -0.194 \\ 
  & (0.523) & (0.526) & (0.523) & (0.521) \\ 
  Income & 0.876$^{***}$ & 0.835$^{***}$ & 0.742$^{**}$ & 0.725$^{**}$ \\ 
  & (0.286) & (0.296) & (0.304) & (0.303) \\ 
  Employed & 1.270 & 1.256 & 1.345 & 1.350 \\ 
  & (0.941) & (0.945) & (0.942) & (0.942) \\ 
  Retiree & 0.893 & 0.889 & 0.770 & 0.811 \\ 
  & (1.380) & (1.387) & (1.382) & (1.381) \\ 
  Unemployed & -5.313$^{*}$ & -5.297$^{*}$ & -5.625$^{**}$ & -5.572$^{**}$ \\ 
  & (2.710) & (2.739) & (2.718) & (2.717) \\ 
  Constant & 12.890$^{***}$ & 12.526$^{***}$ & 13.522$^{***}$ & 13.040$^{***}$ \\ 
  & (2.155) & (2.188) & (2.208) & (2.151) \\ 
 \hline \\[-1.8ex] 
Observations & \multicolumn{1}{c}{299} & \multicolumn{1}{c}{299} & \multicolumn{1}{c}{299} & \multicolumn{1}{c}{299} \\ 
R$^{2}$ & \multicolumn{1}{c}{0.110} & \multicolumn{1}{c}{0.106} & \multicolumn{1}{c}{0.115} & \multicolumn{1}{c}{0.112} \\ 
Adjusted R$^{2}$ & \multicolumn{1}{c}{0.082} & \multicolumn{1}{c}{0.075} & \multicolumn{1}{c}{0.084} & \multicolumn{1}{c}{0.084} \\ 
Residual Std. Error & \multicolumn{1}{c}{6.449 (df = 289)} & \multicolumn{1}{c}{6.472 (df = 288)} & \multicolumn{1}{c}{6.442 (df = 288)} & \multicolumn{1}{c}{6.441 (df = 289)} \\ 
F Statistic & \multicolumn{1}{c}{3.963$^{***}$ } & \multicolumn{1}{c}{3.432$^{***}$ } & \multicolumn{1}{c}{3.740$^{***}$ } & \multicolumn{1}{c}{4.053$^{***}$ } \\ 
 & \multicolumn{1}{c}{ (df = 9; 289)} & \multicolumn{1}{c}{(df = 10; 288)} & \multicolumn{1}{c}{ (df = 10; 288)} & \multicolumn{1}{c}{ (df = 9; 289)} \\ 
\hline 
\hline \\[-1.8ex] 
\textit{Note:}  & \multicolumn{4}{r}{$^{*}$p$<$0.1; $^{**}$p$<$0.05; $^{***}$p$<$0.01} \\ 
\end{tabular} 
}
  \caption[Realized losses on risk behavior]{Realized losses on risk behavior} 
  \label{tab:realizedLossesRiskIndex} 
\end{table} 

\end{subappendices}

\printbibliography  
  
\appendix

\end{document}